\documentclass{JHEP3}
\usepackage{amsmath}
\usepackage{axodraw}
\usepackage{cite}
\usepackage{epsfig}
\usepackage{xspace}
\usepackage{graphics}
\usepackage[utf8]{inputenc}
\usepackage{relsize}
\usepackage{ifthen}
\usepackage{dcolumn}

\def\showcommentsflag{0}
\newcommand{\showcomments}{\def\showcommentsflag{1}}

\newcounter{commentcounter}%
\newcommand{\comment}[1]{\ifnum\showcommentsflag > 0%
\addtocounter{commentcounter}{1}%
\Red{\ensuremath{\ddagger^{\arabic{commentcounter}}}}%
\marginpar{\raggedright\tiny\it\Red{\ensuremath{\ddagger^{\arabic{commentcounter}}} #1}}
\fi%
}
\newcommand{\commentdel}[2]{\ifnum\showcommentsflag > 0%
\Red{\sout{#1}}\comment{#2}%
\fi
}
\newcommand{\commentadd}[2]{\ifnum\showcommentsflag > 0%
\comment{#2}\Red{#1}%
\else
#1
\fi
}
\newcommand{\commentchange}[3]{\ifnum\showcommentsflag > 0%
\Red{\sout{#2}}\comment{#3}\Red{#1}%
\else
#1
\fi
}
\newcommand{\nocomment}[1]{\ifnum\showcommentsflag > 0%
{\tiny\it\Red{\{#1}\}}
\fi%
}
\newcommand{\nocommentdel}[1]{\ifnum\showcommentsflag > 0%
\Red{\sout{#1}}%
\fi
}
\newcommand{\nocommentadd}[1]{\ifnum\showcommentsflag > 0%
\Red{#1}%
\else
#1
\fi
}
\newcommand{\nocommentchange}[2]{\ifnum\showcommentsflag > 0%
\Red{\sout{#2}}\Red{#1}%
\else
#1
\fi
}

\providecommand{\eqref}[1]{eq.~(\ref{#1})\xspace}
\newcommand{\eq}[1]{(\ref{#1})\xspace}
\renewcommand{\eqref}[1]{eq.~(\ref{#1})\xspace}

\newcommand{\eqsref}[1]{eqs.~(\ref{#1})\xspace}

\newcommand{\fig}[1]{\ref{#1}}
\newcommand{\fignum}[1]{figure~#1}
\newcommand{\figref}[1]{figure~\fig{#1}}
\newcommand{\figrefs}[1]{figures~\fig{#1}}
\newcommand{\Figref}[1]{Figure~\fig{#1}}

\newcommand{\sect}[1]{\ref{#1}}
\newcommand{\sectref}[1]{section~\sect{#1}}
\newcommand{\sectrefs}[1]{sections~\sect{#1}}
\newcommand{\Sectref}[1]{Section~\sect{#1}}

\newcommand{\appref}[1]{appendix~\sect{#1}}

\newcommand{\pythia}{P{\smaller YTHIA}\xspace}

\newcommand{\herwig}{H{\smaller ERWIG}\xspace}
\newcommand{\dipsy}{{\smaller DIPSY}\xspace}

\def\done#1{}
\def\pmb#1{{\mbox{\boldmath$#1$}}}
\def\text{\mathrm}
\def\eg{\emph{e.g.}}
\def\ie{\emph{i.e.}}
\def\cf{\emph{c.f.}}

\def\mrm#1{\mathrm{#1}}
\def\sub#1{\ensuremath{_{\mrm{#1}}}}
\def\sup#1{\ensuremath{^{\mrm{#1}}}}
\def\subinelnd{\sub{in,ND}}
\def\subineltot{\sub{in}}
\def\subel{\sub{el}}
\def\subqel{\sub{el*}}
\def\subtot{\sub{tot}}
\def\subsd{\sub{SD}}

\def\subsda{\sub{SD,A}}
\def\subsdp{\sub{SD,p}}

\def\subdd{\sub{DD}}
\def\subd{\sub{D}}

\def\pp{\ensuremath{\mrm{pp}}}
\def\pA{\ensuremath{\mrm{p}A}}
\def\pPb{\ensuremath{\mrm{pPb}}}

\def\pO{\ensuremath{\mrm{pO}}}
\def\gA{\ensuremath{\gamma^\star A}}
\def\gN{\ensuremath{\gamma^\star N}}
\def\gp{\ensuremath{\gamma^\star\mrm{p}}}

\def\gAu{\ensuremath{\gamma^\star\mrm{Au}}}
\def\gCu{\ensuremath{\gamma^\star\mrm{Cu}}}
\def\gO{\ensuremath{\gamma^\star\mrm{O}}}
\def\eA{\ensuremath{\mrm{e}A}}

\def\ep{\ensuremath{\mrm{ep}}}
\def\AA{\ensuremath{AA}}
\def\NN{\ensuremath{NN}}
\def\dAu{\ensuremath{\mrm{dAu}}}
\def\dA{\ensuremath{\mrm{d}A}}
\def\AuAu{\ensuremath{\mrm{AuAu}}}
\def\PbPb{\ensuremath{\mrm{PbPb}}}
\def\Au{\ensuremath{\mrm{Au}}}
\def\Pb{\ensuremath{\mrm{Pb}}}
\def\Cu{\ensuremath{\mrm{Cu}}}

\showcomments
\preprint{ arXiv:1506.09095 [hep-ph] \\
LU-TP 15-24\\
MCnet-15-14\\
\today}

\skip\footins = 1\bigskipamount plus 2pt minus 4pt                              

\title{\boldmath Total, inelastic and (quasi-)elastic cross sections
  of high energy \pA\ and \gA\ reactions with the dipole formalism
  \footnote{Work supported in parts by the MCnetITN FP7 Marie Curie
    Initial Training Network, contract PITN-GA-2012-315877; the
    Swedish Research Council, contracts 621-2012-2283 and
    621-2013-4287; and by the Hungarian OTKA grant NK-101438.}}

\author{G\"osta Gustafson$^1$,  Leif L\"onnblad$^{1}$, Andr{\'a}s Ster$^{1,2}$, and Tam\'as Cs\"org\H{o}$^{2,3}$\\
  $^1$Dept.~of Astronomy and Theoretical Physics, Lund University, Sweden\\
 $^2$MTA Wigner FK, RMI, H-1525 Budapest 114, POBox 49, Hungary\\
 $^3$KRF, H-3200 Gy\"ongy\"os, M\'{a}trai \'{u}t 36, Hungary}
  
\abstract{

  In order to understand the initial partonic state in proton--nucleus
  and electron--nucleus collisions, we investigate the total,
  inelastic, and (quasi-)elastic cross sections in \pA\ and \gA\
  collisions, as these observables are insensitive to possible
  collective effects in the final state interactions. We used as a
  tool the DIPSY dipole model, which is based on BFKL dynamics
  including non-leading effects, saturation, and colour interference,
  which we have extended to describe collisions of protons and virtual
  photons with nuclei. We present results for collisions with O, Cu,
  and Pb nuclei, and reproduce preliminary data on the pPb inelastic
  cross section at LHC by CMS and LHCb. The large \NN\ cross section
  results in \pA\ scattering that scales approximately with the
  area. The results are compared with conventional Glauber model
  calculations, and we note that the more subtle dynamical effects are
  more easily studied in the ratios between the total, inelastic and
  (quasi-)elastic cross sections. The smaller photon interaction makes
  the \gA\ collisions more closely proportional to $A$, and we see here
  that future electron--ion colliders would be valuable complements to
  the \pA\ collisions in studies of dynamical effects from correlations,
  coherence and fluctuations in the initial state in high energy
  nuclear collisions.

 }

\begin{document}
 
\sloppy

\section{Introduction}
\label{sec:intro}

An important question in the understanding of the strong force is the
behaviour of a hot and dense plasma, and properties of the QCD phase
diagram, with a possible critical point.  We note that it is here very
difficult to get definitive answers from ab initio lattice
simulations~\cite{Fodor:2001pe,Fodor:2004nz}.  Results from RHIC,
summarised in the ``White Papers'' by the four
collaborations~\cite{Arsene:2004fa,
  Adcox:2004mh,Back:2004je,Adams:2005dq}, have been interpreted as
indicating the dynamics of a nearly perfect fluid. These results were
subsequently extended to higher energies and investigated in a broader
kinematic range with improved precision by the LHC experiments that
have heavy ion programmes.  These experiments include ALICE, ATLAS and
CMS, and their status during 2014 were summarised for example by
references~\cite{Grosse-Oetringhaus:2014sga,
  AlexanderMilovonbehalfoftheATLAS:2014rta,
  GranierdeCassagnac:2014jha}. The \pPb\ program at the LHC turned out
to be very surprising, culminating in observation of signals of
collectivity not only in \PbPb, but also in \pPb\ reactions.  Recent
results from the Beam Energy Scan program at RHIC~\cite{Adare:2014qvs}
are adding important elements to the picture in the energy range where
many observables change together. These observations have also been
interpreted as a signal for a critical point in the theory of strong
interactions, QCD, in the vicinity of $(T_E,\mu_{B,E}) = ( 165,
95)$~MeV~\cite{Lacey:2014wqa}.

It has often been suggested that studies of \pA\ or \eA\ scattering
can play an important r\^{o}le in studies of the transition between
the ``simpler'' \pp\ or \ep\ collisions and the more complicated \AA\
collisions.  A recent review~\cite{Sickles:2013dca} on \dAu\ and \pPb\
collisions at RHIC and LHC, respectively, indicates that many of the
signatures of nearly perfect fluid dynamics considered for \AuAu\ or
\PbPb\ collisions appear also in \dA\ and \pA\ collisions at RHIC and
LHC energies, thus suggesting that similar perfectly flowing quark
matter may be created also in these collisions, although its volume
and its lifetime must be smaller than in the corresponding heavy ion
collisions. The similarity between \pPb\ and \PbPb\ collisions is also
emphasised in the CMS overview in
ref.~\cite{GranierdeCassagnac:2014jha}.

We also note that even high energy \pp\ collisions exhibit many
features interpreted as signals for collective behaviour, such as
systematics of single-particle spectra and Bose--Einstein correlation
radii~\cite{Agababyan:1997wd}, the increased production of strange
particles and baryons (not least of strange baryons)
\cite{Alner:1985ra,Khachatryan:2011tm,Adams:2006nd}, and signatures
expected from hydrodynamic flow like the p$/\pi$ ratio
\cite{Ortiz:2013yxa} and a near side
''ridge''~\cite{Khachatryan:2010gv}.

Model predictions for final states in \AA\ or \pA\ collisions depend
strongly on the initial conditions obtained from parton--parton
sub-collisions.  In many analyses the initial conditions are estimated
using the Glauber model~\cite{Glauber:1955qq,Glauber:1970jm}. Based on
multiple diffractive sub-collisions, it automatically satisfies
unitarity.  It was early pointed out by Gribov~\cite{Gribov:1968jf},
that diffractive excitation of the intermediate nucleons gives a
significant contribution (on the 10-15\% level) to Glauber's original
model.  Within the Good--Walker formalism \cite{Good:1960ba}
diffractive excitation can be interpreted as a result of fluctuations
in the nucleon's partonic substructure \cite{Miettinen:1978jb}.  It
was also pointed out by Bia\l as, Bleszy\'{n}ski and
Czy\.{z}~\cite{Bialas:1976ed} that fluctuations and correlations in
the positions of the individual nucleons within a nucleus has an
important effect. Such fluctuations also lead to so called quasi-elastic
collisions, where the nucleus is diffractively excited.  These effects
of fluctuations are most easily taken into account in a Monte Carlo
(MC) simulation, where it is also easy to account for a realistic
nucleon distribution within the nucleus, including correlations from a
hard core in the \NN\ interaction (see \eg\
refs.~\cite{Broniowski:2007nz,Rybczynski:2013yba,Alvioli:2009ab}).
 
However, based on independent nucleon--nucleon collisions the Glauber
model does not naturally include possible interference effects between
partons in different nucleons, and it is also difficult to include
uncorrelated fluctuations between the nucleons in a target nucleus.
We also note that although Monte Carlos including diffractive
excitation are available, see \eg\ ref.~\cite{Alvioli:2013vk} and
further references therein, lacking good experimental data for
inelastic diffraction, these corrections are frequently neglected, and
the calculations and implementations of the Glauber model are
typically based on either the total \NN\ cross section or the
inelastic cross section.

The high parton densities at high energy imply, however, that
saturation and interference effects ought to be important. To draw
firm conclusions about a collective behaviour in the final state
interaction, it is therefore crucial to be able to separate features
of the initial state from the dynamics of final state
interactions. The aim of this paper is to investigate the initial
state of a colliding nucleus in greater detail, and to isolate it from
effects of final state interactions. To this end we focus on total,
elastic, and quasi-elastic cross sections in \pA\ and \eA\ collisions,
which are insensitive to effects of final state interactions.

Going beyond the standard Monte Carlo implementations of the Glauber
model, we are in particular interested in the following features:
\begin{itemize}\itemsep 0mm
\item Effects of the partonic substructure in individual nucleons
  within a nucleus.
\item Interference between partons located in different nucleons, but
  having the same colour.
\item Diffractive excitation including high mass diffraction and
  quasi-elastic scattering.
\end{itemize}

Saturation and correlation effects are most easily treated in
transverse coordinate space.  For our analysis we will use the
so-called \dipsy model
\cite{Avsar:2005iz,Avsar:2006jy,Avsar:2007xg,Flensburg:2011kk}, which
is based on BFKL evolution but includes essential non-leading-log
effects, confinement, and saturation within the evolution. \dipsy was
originally developed for \gp\ and \pp\ collisions, but can be directly
extended to model collisions with nuclei~\cite{Flensburg:2012zz}.
Besides making it easier to include correlations, fluctuations, and
diffraction, the formulation in transverse coordinate space also makes
the extension to model realistic nuclei fairly
straight-forward. However, as the gluons from different nucleons can
interfere, the individual nucleons are not quite independent, and we
therefore expect non-trivial effects of shadowing.

The outline of the paper is as follows. First, we will in
\sectref{sec:dipsy} briefly go through the theoretical foundation of
the dipole formalism and its implementation in the \dipsy Monte
Carlo. \Sectref{sec:nuclei} deals with the modifications introduced in
\dipsy to simulate collisions including heavy ions, and the way we
distribute the nucleons inside a nucleus. In \sectref{sec:results} we
first demonstrate that \dipsy reproduces total \pp\ and preliminary \pA\
cross sections from RHIC to LHC energies ($\sqrt{s_{(NN)}} = 200$~GeV
to 8~TeV), before we describe our results and predictions for the
total, inelastic, and (quasi-)elastic \pA\ and \gA\ cross sections.
In \sectref{sec:effects} we discuss the interpretation of the results,
and end the paper with a conclusion and outlook in
\sectref{sec:conclusion}.

\section{The Lund Dipole Cascade Model \dipsy}
\label{sec:dipsy}

Most event generator models of hadronic interactions at high energies
are based on multi-parton interactions, some examples are
\pythia~\cite{Sjostrand:2014zea}, \herwig~\cite{Bahr:2008pv},
DTUJET~\cite{Aurenche:1994ev} and EPOS-LHC~\cite{Pierog:2013ria}.  The
strongly rising parton distributions for small $x$, consistent with
BFKL evolution (albeit with important non-leading effects), make the
parton--parton cross sections very large. At higher energies,
non-linear effects therefore become increasingly important, and these
effects are further enhanced in collisions with nuclei. The non-linear
effects depend on the extension of the cascades in transverse space
(see the pioneering work by Gribov, Levin, and Ryskin
(GLR)~\cite{Gribov:1984tu}). While models for hadronic collisions
generally are formulated in momentum space, analyses of saturation and
non-linear effects are therefore more easily treated in transverse
coordinate space and rapidity.

Formalisms for interactions with high density targets often assume a
homogenous target with no boundary effects (which facilitates the
calculation of analytic results), and many applications are based on
the Color Glass Condensate (CGC) \cite{McLerran:1993ni} or the BK
equation \cite{Balitsky:1995ub,Kovchegov:1999yj}. Here also effects of
correlations and fluctuations are frequently neglected.

The \dipsy model was first developed for hadronic collisions and DIS,
but could also be directly applied to collisions with
nuclei~\cite{Flensburg:2008ag}.  Implemented in the form of a Monte
Carlo event generator, it can easily take into account effects of
correlations, fluctuations, and finite nuclear geometry. When
modelling \pA\ or $AB$ nuclear reactions with \dipsy it is assumed
that these interactions are dominated by absorption into inelastic
channels. Thus the imaginary part of the $S$-matrix is neglected, and
the total, elastic, and quasi-elastic cross sections are directly
obtained via the optical theorem. The assumed reality of the
$S$-matrix also implies that it is possible to take the Fourier
transform of the amplitude in $b$-space. This method was applied to
obtain the differential elastic cross section $d\sigma/dt$ of \pp\
collisions from \dipsy in ref.~\cite{Flensburg:2008ag}.

\subsection{Dipole evolution in transverse coordinate space}
\label{sec:dipsy-model}

The \dipsy model is based on Mueller's dipole cascade model
\cite{Mueller:1993rr,Mueller:1994jq,Mueller:1994gb}, which is a
formulation of leading-log BFKL
evolution\cite{Kuraev:1977fs,Balitsky:1978ic} in transverse coordinate
space. Mueller's model relies on the fact that initial-state radiation
from a colour charge (in a quark or a gluon) within a hadron is
screened at large transverse distances by an accompanying anti-charge,
and that gluon emissions therefore can be described in terms of
colour-dipole radiation. Thus the partonic state is described in terms
of dipoles in impact-parameter space, evolved in rapidity when a
dipole is split into two dipoles by gluon emission.  The screening
implies a suppression of large dipoles in transverse coordinate space,
which is equivalent to the suppression of small $k_\perp$ in the
conventional BFKL evolution in momentum space.

For a dipole with charges at the transverse points $\mathbf{x}$ and
$\mathbf{y}$, the probability per unit rapidity ($Y$) to emit a gluon
at $\mathbf{z}$ is given by
\begin{equation}
  \label{eq:dipoleradiation}
  \frac{d\mathcal{P}_g}{dY} = \frac{\bar{\alpha}}{2\pi}d^2\pmb{z}
  \frac{(\pmb{x}-\pmb{y})^2}{(\pmb{x}-\pmb{z})^2 (\pmb{z}-\pmb{y})^2}\,,
  \,\,\,\,\,\,\, \mathrm{with}\,\,\, 
  \bar{\alpha} = \frac{N_c\alpha_s}{\pi}.
\end{equation}
The emission produces two new dipoles, $(\pmb{x},\pmb{z})$ and
$(\pmb{z},\pmb{y})$, which can split independently by further gluon
emissions. Repeated emissions form a cascade, with dipoles connected
in a chain.  The chain can be thought of as dipoles connected by
gluons, or gluons connected by dipoles~\cite{Gustafson:2002kz}. When
two cascades collide, a dipole $(\pmb{x}_1,\pmb{y}_1)$ in a
right-moving cascade can interact with a left-moving dipole
$(\pmb{x}_2,\pmb{y}_2)$, with probability (in Born approximation)
\begin{equation}
  \label{eq:dipoleinteraction}
  P_{12}=\frac{\alpha_s^2}{4}
  \left[\ln\left( \frac{(\pmb{x}_1-\pmb{y}_2)^2(\pmb{y}_1-\pmb{x}_2)^2} 
      {(\pmb{x}_1-\pmb{x}_2)^2(\pmb{y}_1-\pmb{y}_2)^2}\right)\right]^2.
\end{equation}

We note that in this leading log approximation, the probability for
dipole emission in \eqref{eq:dipoleradiation} diverges for small
dipole sizes $(\pmb{x}-\pmb{z})$ or $(\pmb{z}-\pmb{y})$, but as the
interaction probability in \eqref{eq:dipoleinteraction} then goes to
zero, the total cross section is finite. (This is naturally related to
colour transparency.) Although the final result is finite, the
divergence causes numerical problems in a leading log Monte Carlo
simulation \cite{Salam:1996nb}.  This problem is removed by the
non-leading effects included in \dipsy, which suppress very small
dipoles.

Via the optical theorem the expression in \eqref{eq:dipoleinteraction}
also equals twice the elastic dipole--dipole scattering amplitude. (As
mentioned above we assume that the interaction is driven by the
inelastic absorption.) Denoting the right-moving dipoles $i$ and the
left-moving $j$, the total elastic scattering amplitude is in the Born
approximation given by $F=\sum_{ij} P_{ij}/2$. Multiple scatterings
are taken into account in the eikonal approximation via the unitarized
elastic amplitude (defined by the relation $S \equiv 1-T$, which makes
$T$ real)
\begin{equation} 
  T=1-\exp{(-F)}=1-\exp{(-\sum P_{ij}/2)}.  \label{eq:unitaryamplitude} 
\end{equation}
To get the full elastic amplitude we also have take the average over
all possible cascades.  In impact parameter space, the differential
total and elastic cross sections are then, via the optical theorem,
given by
\begin{eqnarray}
  d\sigma\subtot/d^2 b &=& 2\langle T \rangle, \nonumber \\
  d\sigma\subel/d^2 b &=& \langle T \rangle^2.
\label{eq:crossections}
\end{eqnarray}

The cross section for diffractive excitation can be obtained in the
Good--Walker \cite{Good:1960ba} formalism, which is described in
\appref{sec:appendix}. (Diffractive excitation is usually described in
either the Good--Walker or the triple-reggeon formalism, but as
demonstrated in ref.~\cite{Gustafson:2012hg}, for high mass
diffraction these are just different sides of the same BFKL evolution
dynamics.)  Diffractive excitation is here determined by the
fluctuations in the scattering process.  A proton is a linear
combination of all possible parton cascades, and as discussed in
refs.~\cite{Avsar:2007xg,Flensburg:2010kq}, we assume that these
cascades represent the diffractive eigenstates, which thus can come on
shell in the interaction.  The cross section for diffractive
excitation is thus described by the fluctuations among the possible
parton cascades in the projectile and the target.

As described in the Appendix, the total diffractive cross section,
including elastic scattering, is obtained by summing over all possible
diffractive states, and given by $\langle T^2 \rangle$. This is the
sum of elastic scattering, single diffraction of the projectile or the
target, and double diffraction. Single diffraction of the projectile
with an elastic target is similarly obtained by summing over all
projectile states but keeping the target intact, and then subtract the
elastic scattering (which is still given by $\langle T
\rangle^2$). The result is
\begin{equation}
  d\sigma\sub{SD,p}/d^2 b =\langle\langle T \rangle_{t}^2\rangle_p-
  \langle T \rangle_{p,t}^2.
\label{eq:SD}
\end{equation}
The averages are here taken over all possible cascades for the
projectile ($P$) and the target ($T$), for a fixed impact parameter
$b$.  (Note again that the amplitude T, given by
\eqref{eq:unitaryamplitude}, is a real quantity.)

We note that summing over projectile cascades gives production of all
possible cascades limited by the rapidity determined by the Lorentz
frame used in the analysis. This gives also the possibility to
calculate the differential single diffractive cross section
$d\sigma\subsd/dM_X^2$.

Double diffraction is similarly given by total diffraction minus
single diffraction and minus elastic scattering, which gives the
result
\begin{equation}
  d\sigma\subdd/d^2 b =\langle T^2\rangle_{P,T} -
  \langle\langle T \rangle_{T}^2\rangle_P - 
  \langle\langle T \rangle_{P}^2\rangle_T +
  \langle T \rangle_{P,T}^2.
\label{eq:DD}
\end{equation}

\subsection{Beyond leading log}
In a series of papers
\cite{Avsar:2005iz,Avsar:2006jy,Avsar:2007xg,Flensburg:2011kk} a
generalisation of Mueller's model, implemented in the Monte Carlo
event generator \dipsy, has been described in detail. Here we will
only discuss the main points.  The basic idea behind the model is to
include important non-leading effects in the BFKL evolution,
saturation effects within the evolution, and confinement.

The full next-to-leading logarithmic corrections to BFKL have been
calculated and have been found to be very large
\cite{Fadin:1998py,Ciafaloni:1998gs}.  A physical interpretation of
these corrections has been presented by Salam \cite{Salam:1999cn}, and
a dominant part is related to energy-momentum conservation. In the
\dipsy model this is achieved by equating the emission of a gluon at
small transverse distances with high transverse momenta for the
emitted and recoiling gluons. Thus the conservation of energy and
momentum implies a dynamic cutoff for very small dipoles with
correspondingly high transverse momenta. This constraint has also
important computational advantages, removing the divergence in the
dipole splitting probability in \eqref{eq:dipoleradiation}.  Other
important non-leading effects are the running coupling,
$\alpha_s(p_\perp^2)$, and the ``consistency constraint" or ``the
energy scale terms", which implies that the emissions are ordered in
both the positive and negative light-cone components
\cite{Ciafaloni:1987ur,Andersson:1995ju,Kwiecinski:1996td}.

\subsection{Nonlinear effects}
\label{sec:swing}

Besides these perturbative corrections, confinement effects are
included via a small gluon mass, and non-linear saturation effects
through the so-called swing mechanism, which is a colour suppressed
correction to the cascade evolution.  Mueller's dipole evolution is
derived in the large $N_c$ limit, where each colour charge is uniquely
connected to an anti-charge within a dipole.  Saturation effects are
here included as a result of multiple dipole interactions, in the
frame used for the analysis. Such multiple interactions give dipole
chains forming loops (see ref.~\cite{Avsar:2006jy}), and are related
to multiple pomeron exchange. When multiple interactions can be
neglected, Mueller's model gives a result, which is independent of the
Lorentz frame used for the calculation ~\cite{Mueller:1996te}.  Loops
formed within the evolution are, however, not included. This means
that when multiple interactions cannot be neglected, the result is no
longer frame independent. (Loops within the evolution are also
neglected \eg\ in the non-linear BK
equation~\cite{Balitsky:1995ub,Kovchegov:1999yj}.)

The dipole interaction in \eqref{eq:dipoleinteraction} is proportional
to $\alpha_s$, and thus colour suppressed compared to the dipole
splitting in \eqref{eq:dipoleradiation}. Loop formation is therefore
related to the possibility that two dipoles have identical colours.
For two dipoles with the same colour, we have actually a colour
quadrupole, where a colour charge is effectively screened by the
nearest anti-colour charge.  Thus approximating the field by a sum of
two dipoles, one should preferentially combine a colour charge with a
nearby anti-charge.  This interference effect is taken into account in
\dipsy in an approximate way, by allowing two dipoles with the same
colour to recouple forming new dipoles, in a way favouring small
dipoles. This mechanism, called a swing, is illustrated in
\figref{fig:swing}; see further ref.~\cite{Avsar:2006jy}.

\FIGURE[t]{
\centering
\begin{picture}(280,100)(-50,0)
\Line(10,10)(20,25)
\Line(25,40)(20,25)
\Line(50,40)(80,30)
\Line(75,60)(80,30)
\Line(75,60)(55,50)
\Line(30,50)(15,70)
\Line(15,70)(10,90)
\LongArrow(85,45)(100,45)
\LongArrow(100,45)(85,45)
\Line(100,10)(110,25)
\Line(115,40)(110,25)
\Line(140,40)(170,30)
\Line(165,60)(170,30)
\Line(165,60)(145,50)
\Line(120,50)(105,70)
\Line(105,70)(100,90)
\Vertex(10,10){2}
\Vertex(20,25){2}
\Vertex(25,40){2} 
\Vertex(50,40){2} 
\Vertex(80,30){2} 
\Vertex(75,60){2}
\Vertex(55,50){2} \Vertex(30,50){2} \Vertex(15,70){2} \Vertex(10,90){2}
\Vertex(100,10){2} \Vertex(110,25){2} \Vertex(115,40){2} \Vertex(120,50){2}
\Vertex(140,40){2} \Vertex(170,30){2} \Vertex(165,60){2} \Vertex(145,50){2}
\Vertex(105,70){2} \Vertex(100,90){2}
\SetColor{Red}
\Line(25,40)(50,40)
\Line(30,50)(55,50)
\Line(115,40)(120,50)
\Line(145,50)(140,40)
\Text(25,39)[tl]{$\bar{r}$}
\Text(50,37)[tr]{$r$}
\Text(30,53)[bl]{$r$}
\Text(55,52)[br]{$\bar{r}$} 
\end{picture}
\caption{\label{fig:swing}Two dipoles with the same colour, forming a
  colour quadrupole, may be better approximated by two dipoles formed
  when the colour charges are recoupled, as illustrated here.}
}

Although this recipe is an approximation, the dependence on the
Lorentz frame used for calculating the cross section is significantly
reduced, as demonstrated and detailed in ~\sectref{sec:frindep}.

In the simulation the swing is handled by assigning all dipoles a
colour index running from 1 to $N_c^2$, not allowing two dipoles
connected to the same gluon to have the same index.  A pair of two
dipoles, $(\pmb{x}_1,\pmb{y}_1)$ and $(\pmb{x}_2,\pmb{y}_2)$, having
the same colour, may be better approximated by the combination
$(\pmb{x}_1,\pmb{y}_2)$ and $(\pmb{x}_2,\pmb{y}_1)$, if these dipoles
are smaller.  In the evolution the pair is allowed to ``swing" back
and forth between the two possible configurations shown in
\figref{fig:swing}.  The swing mechanism is adjusted to give the
relative probabilities $(\pmb{x}_1-\pmb{y}_2)^2(\pmb{x}_2-\pmb{y}_1)^2
: (\pmb{x}_1-\pmb{y}_1)^2(\pmb{x}_2-\pmb{y}_2)^2$, thus favouring the
configuration with smallest dipoles.  In the evolution in rapidity,
the swing is competing with the gluon emission in
\eqref{eq:dipoleradiation}, where a Sudakov-veto algorithm
\cite{Sjostrand:2006za} can be used to choose which of the two happens
first.

The non-linear GLR~\cite{Gribov:1984tu} and
BK~\cite{Balitsky:1995ub,Kovchegov:1999yj} equations are formulated in
a way where the number of dipoles can be reduced, via a $(2\rightarrow
1)$ vertex. What is calculated in these formalisms is actually the
interaction probability. With the swing mechanism the number of
dipoles is unchanged. However, as the dipole swing leads to smaller
average dipole sizes, the interaction probability in
\eqref{eq:dipoleinteraction} is also reduced in accordance with colour
transparency. Thus the swing represents a similar kind of physics as
these evolution equations. An advantage with the \dipsy formalism is
that it is possible to also take into account
correlations~\cite{Flensburg:2011kj}, fluctuations, and effects of the
nuclear size and shape.

\subsection{Final states}
\label{sec:finalstates}

The BFKL equation and Mueller's model describe the inclusive cross
section, but as seen from the CCFM formalism
\cite{Ciafaloni:1987ur,Catani:1989yc} it is possible to include softer
gluon emissions to form exclusive final states. As described in
ref.~\cite{Andersson:1995ju} in the Linked Dipole Chain model (LDC),
these softer gluons can be added as final state radiation from an
initial BFKL ladder.  When the two evolved systems collide, some of
the dipoles in the right-moving system interact with some in the
left-moving one, and the cross sections are calculated via
\eqsref{eq:dipoleradiation} - \eq{eq:crossections}.  The interaction
enables the gluons in the interacting dipoles to come on-shell,
together with all parent dipoles, while non-interacting dipoles must
be regarded as virtual and thus be reabsorbed. In a situation, where
saturation is important, it is also necessary to include possible
swings in the final state~\cite{Avsar:2006jy,Flensburg:2012zy}.  In
the \dipsy MC the gluons are traced in both momentum and coordinate
space, and therefore the model also allows us to generate the
final-state momentum distribution of gluons and, after hadronization,
final hadronic states~\cite{Flensburg:2011kk,Flensburg:2012zy}.  In
the present paper we restrict our study to inclusive total and
inelastic cross sections, while final states will be studied in future
work on \pA\ and \eA\ collisions.

The stochastic nature of BFKL evolution also implies, that the
exclusive final states obtained in an event generator like \dipsy,
includes fluctuations and correlations which are essential for many
observables. As an example, we note that the fluctuations in the
scattering amplitude determines the cross section for diffractive
excitation in \eqref{eq:crossections}.

\subsection{The Monte Carlo event generator \dipsy}
\label{sec:application}

The Lund Dipole Cascade Model has been implemented in the Monte Carlo
event generator called \dipsy, with applications mainly to \pp\
collisions and DIS.  In a high energy collision, two hadrons are
evolved from their respective rest frames to a Lorentz frame in which
they collide.  In its own rest system a proton is currently modelled by
a simple triangle of gluons connected by dipoles, and the gluonic Fock
state is built by successive dipole emissions of virtual gluons. The
small-$x$ partons are dominated by gluons, and the result for small
$x$ is rather insensitive to the choice of initial parton
configuration, apart from its overall size.  We also note that the
proton structure functions are well reproduced by the \dipsy model
\cite{Avsar:2007ht}. (Quarks are important for the final state, and
here valence quarks are later introduced by hand).

The parameters of the MC model have been tuned to the energy
dependence of the total and elastic \pp\ scattering cross sections.  The
best values of the tune parameters that govern the calculations will
be shown and discussed in \sectref{sec:pp-tune}. Since the system at
very large energies is dominated by gluons, the model may use the same
parameters later on for all kinds of reactions and observables.

Here, we show the list of the parameters whose values substantially
influence the final results.  They are described in more detail for
example in ref.~\cite{Flensburg:2008ag}.

\begin{itemize}\itemsep 0mm
\item $R_{max}$: Non-perturbative regularisation scale, this
  corresponds to the maximum dipole size in a given simulation, above
  which emissions and interactions are exponentially suppressed
  \cite{Avsar:2007xg}. Its typical value is $\approx3$~GeV$^{-1}$.
\item $R_{p}$: The average size of the proton at rest. Its typical
  value is $\approx3$~GeV$^{-1}$.
\item $w_{p}$: The width of the Gaussian fluctuations in proton size
  around $R_p$. As in ref.~\cite{Flensburg:2008ag} this value was set
  to 0.1~GeV$^{-1}$.
\item $\Lambda_{QCD}$: This is the scale parameter of $\alpha_{s}$,
  the running coupling constant of QCD, and its default value in the
  present calculations was 0.23~GeV.
\item $\lambda_{r}$: This parameter controls the swing effect, its
  default value in the presented calculations was 1. Dipoles with the
  same colour are allowed to swing back and forth, which results in an
  equilibrium, where the smaller dipoles have a larger
  weight. $\lambda_{r}=1$ has previously \cite{Flensburg:2011kk} been
  shown to be large enough to reach the equilibrium.
\end{itemize}

\section{Treatment of nuclei in \dipsy}
\label{sec:nuclei}

The \dipsy model, initially developed for \ep\ and \pp\ collisions,
can be directly generalised to simulate collisions with nuclei. We
just have to generate random positions for all the nucleons within a
target nucleus, and let them collide with a projectile proton, a
virtual photon, or a similarly generated projectile nucleus. In the
present paper we will focus on high energy \pA\ and \gA\ scattering,
and of particular interest will here be effects of colour interference
between different nucleons in a nucleus, and effects of fluctuations,
both in the distribution of nucleons in a nucleus, and in the parton
evolution inside individual nucleons.

High energy nuclear collisions are usually analysed within the Glauber
formalism~\cite{Miller:2007ri,Glauber:1955qq}. Here it is assumed that
the projectile nucleon(s) travel along straight lines and undergo
multiple diffractive sub-collisions.  Inspite of its pure geometric
approach, it has been quite successful in describing many
characteristics of reactions with nuclei, and has been widely used in
experiments at RHIC and LHC, \eg\ to estimate the number of binary
nucleon--nucleon collisions and the number of participant nucleons as
a function of centrality.  In order to illustrate some features of our
results, we will below also make comparisons with Glauber simulations.

Concerning fluctuations we note that there are two different origins
for fluctuations i collisions with nuclei.  The first is due to
fluctuations in the position of the nucleons in a nucleus, while the
second is related to diffractive excitation of the wounded
nucleons. As described in \sectref{sec:dipsy-model}, diffractive
excitation of a proton is determined by fluctuations in the internal
proton substructure, as given by the Good--Walker formalism. The two
effects, and their relation to the Glauber formalism, will be
described in \sectrefs{sec:nucleongliss} and \sect{sec:diff}. The
Good--Walker formalism can also be directly applied to calculate
quasi-elastic scattering, where the nucleus is breaking up. This is
also discussed in \sectref{sec:diff}.

A very interesting question is also to what extent colour charges in
one nucleon can screen charges in other nucleons. In \dipsy this
interference effect is described by the swing mechanism, discussed in
\sectref{sec:swing}, and in a nucleus we also allow dipoles in
different nucleons to swing. The result is a kind of colour
reconnection, and in the analysis of our results we will compare the
outcome with and without this inter-nucleon swing.  Such possible
interference effects between the different nucleons are not taken into
account in the Glauber approach.

\subsection{Distribution of nucleons in a nucleus}
\label{sec:nucleongliss}

It was early pointed out by Bia\l as, Bleszy\'{n}ski and
Czy\.{z}~\cite{Bialas:1976ed} that fluctuations and correlations in
the positions of the individual nucleons within a nucleus has an
important effect. This effect, which is most easily taken into account
using MC simulation techniques (not easily available at the time) has
been studied by several authors, see \eg\
refs.~\cite{Alvioli:2009ab,Blaizot:2014wba}.  The nucleon positions
are typically generated so as to reproduce the charge density observed
in \eA\ scattering, and the hard core in the \NN\ interaction has an
important effect reducing the fluctuations in the distributions.

\emph{The GLISSANDO method}

We will in our simulations follow the method in the GLISSANDO
MC~\cite{Broniowski:2007nz,Rybczynski:2013yba}, to generate the
nucleon distributions. The nucleon positions are generated according
to a Wood--Saxon distribution for the nucleons, with the following
form:
\begin{equation}
  \rho(r) = 
  \frac{\rho_0 (1 + wr^2/R^2)}
  {1+\exp((r-R)/a)}.
\label{eq:woodsaxon1}
\end{equation}
Here $R$ is the nuclear radius, $a$ is ``skin width", and $\rho_{0}$ is the
central density. The parameter $w$ describes a possible non-constant density, 
but is zero in the fits to nuclei used in this paper.

The nucleon centres are randomly generated in such a way that the
charge distribution determined in ref.~\cite{DeVries:1987qc} is
recovered, when the result is convoluted with the charge distribution
within nucleons. The nucleons are also generated with a hard-core,
which thus introduces short range correlations among the nucleons. As
shown by Rybczynski and Broniowski~\cite{Rybczynski:2010ad}, the
correct two-particle correlation can be obtained if the nucleons are
generated with a minimum distance equal to $2r_{core}$.

We note that if the nucleons are generated within a specific volume,
the resulting distribution will (for a finite nucleus) be confined
within a smaller volume, and its centre will be shifted. According to
ref.~\cite{Rybczynski:2013yba}, the correct charge distribution is,
for mass numbers $A> 16$, obtained using randomly generated nucleon
centres described by the Wood--Saxon form in \eqref{eq:woodsaxon1}
with the following parameters:
\begin{equation}
  R_{NC} =  (1.1A^{1/3} - 0.656A^{-1/3} ) \,\,{\rm fm}, \,\,\,\,a = 0.459
  \,\, {\rm fm}, \,\,\,\, w=0,
\label{eq:gliss_R}
\end{equation}
together with a hard core with radius $r_{core}=0.45$~fm. 

Currently the \dipsy MC includes parametrisations for He, O, Cu, Au,
and Pb\footnote{Other nuclei can easily be added by the user.}.  (Thus
we use the same spherical form also for the light nuclei He and O.)
For the nuclei studied in this paper this corresponds to the following
radii: $R^{\Pb}_{NC}=6.406$~fm, $R^{\Au}_{NC}=6.288$~fm,
$R^{\Cu}_{NC}=4.236$~fm, and $R^{\mrm{O}}_{NC}=2.511$~fm.

\subsection{Diffractive excitation of wounded nucleons}
\label{sec:diff}

In Glauber's initial formulation the nucleus was described by a smooth
distribution determining the absorption probability, but it was early
pointed out by Gribov~\cite{Gribov:1968jf}, that effects of
diffractive excitation of the wounded nucleons are quite significant.
Diffractive excitation occurs when the projectile is a linear
combination of states with different absorption
probabilities. Diffractive excitation is here determined by the
fluctuations in the scattering amplitude, as seen in
\eqref{eq:diffraction}.  At lower energies this could be well
approximated by a single diffractive state, but it is now well known
that diffractive excitation in \pp\ collisions or DIS is not limited
to low masses, and has a high cross section $\sim 10-15\%$ of the
total~\cite{Abe:1993wu,Adloff:1997sc,Chekanov:2005vv}.  An important
point is here that, due to time dilation, in a \pA\ collision the
state of the projectile is frozen between the sub-collisions, while the
target nucleons may all be in different states~\cite{Blaettel:1993ah}.

As mentioned in the introduction, this effect has been studied in
several analyses and, as an example, a model by Strikman and
coworkers~\cite{Guzey:2005tk,Alvioli:2013vk} has been applied by the
\textsc{Atlas} collaboration \cite{TheATLAScollaboration:2013cja}.
However, lacking good experimental data for inelastic diffraction,
these corrections are frequently neglected, with calculations based on
either the total \NN\ cross section or the inelastic cross section
(including or excluding diffraction).

In the simplest form of the Glauber Model, the target in a \NN\
collision acts as a black absorber. The projectile nucleon travels
along a straight line and interacts inelastically if the transverse
distance to a nucleon in the target is smaller than a distance $R$,
with
\begin{equation}
  \pi R^2 = \sigma\subinelnd^{NN}. 
\label{eq:blackdisc}
\end{equation}
Here $\sigma\subinelnd^{NN}$ is the inelastic, non-diffractive
nucleon--nucleon cross section. This black-disc approximation implies
the following cross sections for nucleon--nucleon collisions (see
\eqsref{eq:crossections}, \eq{eq:diffraction}):
\begin{eqnarray}
  \sigma\subtot &=& 2 \int d^2 b \langle T(b) \rangle = 2 \pi R^2 \nonumber \\
  \sigma\subel &=& \int d^2 b \langle T(b) \rangle^2 = \pi R^2 \nonumber \\
  \sigma\subd &=& \int d^2 b (\langle T(b)^2 \rangle - \langle T(b) \rangle^2)
  = 0  \nonumber \\
  \sigma\subinelnd &=& \int d^2 b \langle 1 - (1 - T(b))^2\rangle = \pi R^2.
\label{eq:crossectionsGl}
\end{eqnarray}
These relations demonstrate, that in a black-disc approximation the
diffractive part of the cross section vanishes, and that the elastic
and inelastic cross sections are equal, both given by half the total.

The most simple model accounting for diffractive excitation is the so
called grey disc model. Here it is assumed that within a radius $R$
the projectile is absorbed with probability $a$, with $0<a<1$. The
resulting \pp\ cross sections are here
\begin{eqnarray}
  \sigma\subtot    &=& 2 \int d^2 b \langle T(b) \rangle    = 2 \pi R^2 a                  \nonumber \\
  \sigma\subel     &=&   \int d^2 b \langle T(b) \rangle^2  =   \pi R^2 a^2                 \nonumber \\
  \sigma\subd      &=&   \int d^2 b (\langle T(b)^2 \rangle -
  \langle T(b) \rangle^2) = \pi R^2 a(1-a) \nonumber \\
  \sigma\subinelnd &=&   \int d^2 b \langle 1 - (1 - T(b))^2\rangle        = 
  \pi R^2 a.
\label{eq:greydisc}
\end{eqnarray}
The parameters $R$ and $a$ can now be adjusted to reproduce \eg\ the
total and the elastic \pp\ cross sections.  It is interesting to note
that even in this gray disc limit, $ \sigma\subel + \sigma\subd =
\sigma\subinelnd = \sigma\subtot / 2 .  $ We also note that in the LHC
energy region $\sigma\subel \approx 0.25\, \sigma\subtot$, which in
this grey disc model gives $\sigma\subd \approx \sigma\subel$ and
$\sigma\subinelnd \approx 0.5\, \sigma\subtot$. Also other forms for
the \pp\ interaction have been considered in the literature, \eg\ a
Gaussian interaction profile
\cite{Ding:1989mj,Pi:1992ug,Miller:2007ri}.

In the black disc approximation the fluctuations are \emph{totally
  neglected}. As mentioned above, in a \pA\ collision the state of the
projectile is frozen between the sub-collisions, while the target
nucleons may all be in different states. Averaging over different
colliding target nucleons therefore tends to reduce the
fluctuations. Consequently, if the parameters in the grey disc
approximation are adjusted to the total and elastic \pp\ cross
sections, the effects of fluctuations are instead
\emph{overestimated}.  The averaging over different states for the
target nucleons is properly taken into account in \dipsy, and when
comparing the \dipsy results with Glauber calculations we therefore
expect the \dipsy to lie in between the black and grey disc results
\footnote{This overestimate of the fluctuations in the target is also
  present in other Glauber approaches, where the same diffractive
  eigenstate is used in each \NN\ sub-collision.}.

As the main difference in \pA\ collisions between the \dipsy model and
Glauber-based models is the treatment of fluctuations inside nucleons,
it is interesting to study the diffractive dissociation of the
nucleus. From the single diffraction cross section in \eqref{eq:SD} we
immediately get that this is given by
\begin{equation}
  d\sigma\subsda/d^2b=
  \langle\langle T\rangle_p^2\rangle_A - \langle T\rangle_{p,A}^2\,\,.
  \label{eq:sda}
\end{equation}
Experimentally this means that we must detect an elastically scattered
proton in one direction, and a dissociated nucleus in the other. In
\dipsy there is in this case a dependence on the frame in which the
collision is studied. Typically the proton--nucleon rest frame is
used, which means that we only consider the case where the
diffractively dissociated nucleus does not produce particles in the
proton hemisphere of the detector.

It should be noted that even if the grey- and black-disc Glauber
calculations do not model the single diffractive components for \pp\
collisions, they will give diffractively dissociated nuclei due to the
fluctuations in the distribution of nucleons. This can be thought of
as elastic scattering of the proton with one of the nucleons, causing
the nucleus to break up.

Experimentally it may be difficult to measure low-mass diffraction in
the nucleus direction, and below we will also study what we call
quasi-elastic scattering, where again the proton is scattered
elastically, but the only other requirement is that the proton
hemisphere of the detector is empty. This corresponds to adding the
single diffractive and elastic cross sections, resulting in
\begin{equation}
  d\sigma\subqel/d^2b=
  \langle\langle T\rangle_p^2\rangle_A\,\,.
  \label{eq:qel}
\end{equation}

\section{Monte Carlo cross section results}
\label{sec:results}

In this section we present various Monte Carlo simulation results for
high energy proton--proton, proton--ion, photon--proton and
photon--ion reaction cross sections performed by \dipsy. In order to
show that these numbers are reasonable, they are compared to
(preliminary) data, where these data are already available.  Most of
the (heavy) ion results (\pA\ and \gA) are in fact predictions,
%academic studies
given that the corresponding experimental values 
are not yet measured and many of these determinations 
are not even foreseen at present.

As discussed in \sectref{sec:intro}, our calculations are based on the
evolution and interaction of virtual gluons whose number is increasing
with the increase of $\sqrt{s}$ and the size of the nuclei involved.
As a practical consequence, the simulations for more energetic
reactions and higher values of A are limited by the available
computational power to achieve reasonable statistics in reasonable
time. The scope of this paper, due to this practical consideration,
covers ranges of centre-of-mass energies from the upper range of RHIC
energies of $\sqrt{s_{NN}} = 200$~GeV up to the LHC energies of 8~TeV,
selecting a few typical points in between. The RHIC energy
$\sqrt{s_{NN}} = 200$~GeV for colliding nuclei provides partons with
about $x=0.01$, which is the lowest limit for an acceptable precision
with the event generator.  At lower energies the contribution of the
previously neglected quark degrees of freedom may play quite a
significant role.  The upper range of the energy scale was chosen to
be 8~TeV so that the simulations could be completed with the available
computing power in the timescale of this project.  For the particular
case of the \pPb\ reactions, we have evaluated the total cross section
and its decomposition also for the expected, future LHC energy of
$\sqrt{s_{NN}} = 10$~TeV.

In the figures, as a general rule, \dipsy results are shown for eight different
nucleon--nucleon centre-of-mass energies chosen appropriately from RHIC to LHC
energies in case of \pp\ and \pA\ reactions.  To help the visualisation of the
results, the simulated points were connected by simple straight lines to guide
the eye.  For \gA simulations, one more simulation point was added
at $\sqrt{s_{\gN}} = 100$~GeV for the sake of possible \eA\ experiments
at RHIC where, in the current planning phase, the $\sqrt{s} =  20- 90$~GeV centre
of mass energy range is considered for \eA\ collisions with large $A$,  and the 30
to 145~GeV energy range is considered for polarised \ep\
collisions~\cite{Accardi:2012qut,Burton:2014tqa}.

\subsection{Tuning and comparison to data}
\label{sec:pp-tune}

\subsubsection[\pp\ total and elastic cross sections]
{\boldmath \pp\ total and elastic cross sections}
\label{sec:pp-x}

In \figref{fig:pp}, \dipsy results of pp total and elastic cross
sections are shown after the most sensitive model parameters are tuned
to data.  The ball park values of these parameters are listed in
\sectref{sec:application}, and we give their actual values in this
sub-section.  with the help of these tunes.  Subsequently, these
parameters were fixed and utilised in the more complex simulations of
proton--ion and photon--ion events.  For comparisons they were also
used to simulate photon--proton reactions.

Our plot in \figref{fig:pp} follows the style of \fignum{1} of
ref.~\cite{Antchev:2013paa}, where the TOTEM collaboration compared
its recently measured total and elastic \pp\ cross sections at 7 and
8~TeV with earlier results, and low statistics cosmic ray experiments
at very high energies, see
refs.~\cite{Antchev:2013gaa,Antchev:2013paa} for further details.
Cross sections measured at lower energies were taken from from the
Particle Data Group~\cite{Beringer:1900zz} database.  The best fits
with a formula by the COMPETE Collaboration~\cite{compete:2002} are
also compared to the data on the total and also on the elastic cross
sections with coefficients shown for $\sigma\subel$ in the legend of
the plot.  The results of the simulations by \dipsy (shown by full
dots) indicate that \dipsy simulations follow the trends of the
cross-section data reasonably well.

To achieve this reasonably good level of description of the total and
elastic cross sections, the QCD scale parameter $\Lambda_{QCD}$ in
\dipsy was tuned to the value of 0.230~MeV. The effective size
parameter of the proton was found to be
$R_p=2.9$~GeV$^{-1}\approx0.57$~fm (which is different from 0.45 fm
used for the hard-core radius mentioned above).

The successful reproduction of the experimental data on the total and
elastic cross sections of \pp\ reactions in the energy range of
$\sqrt{s} = 0.2 - 8$~TeV energy range, where \dipsy has valid
approximate assumptions for the parton evolution within the nucleons,
opens the way for Monte Carlo calculations of electron--ion (\eA) and
proton--ion (\pA) reactions. Heavy ion ($AB$) collisions can be
studied in a similar manner, but the investigation of these reactions
goes beyond the scope of this article.

\FIGURE[t]{
  \centering
  \epsfig{file=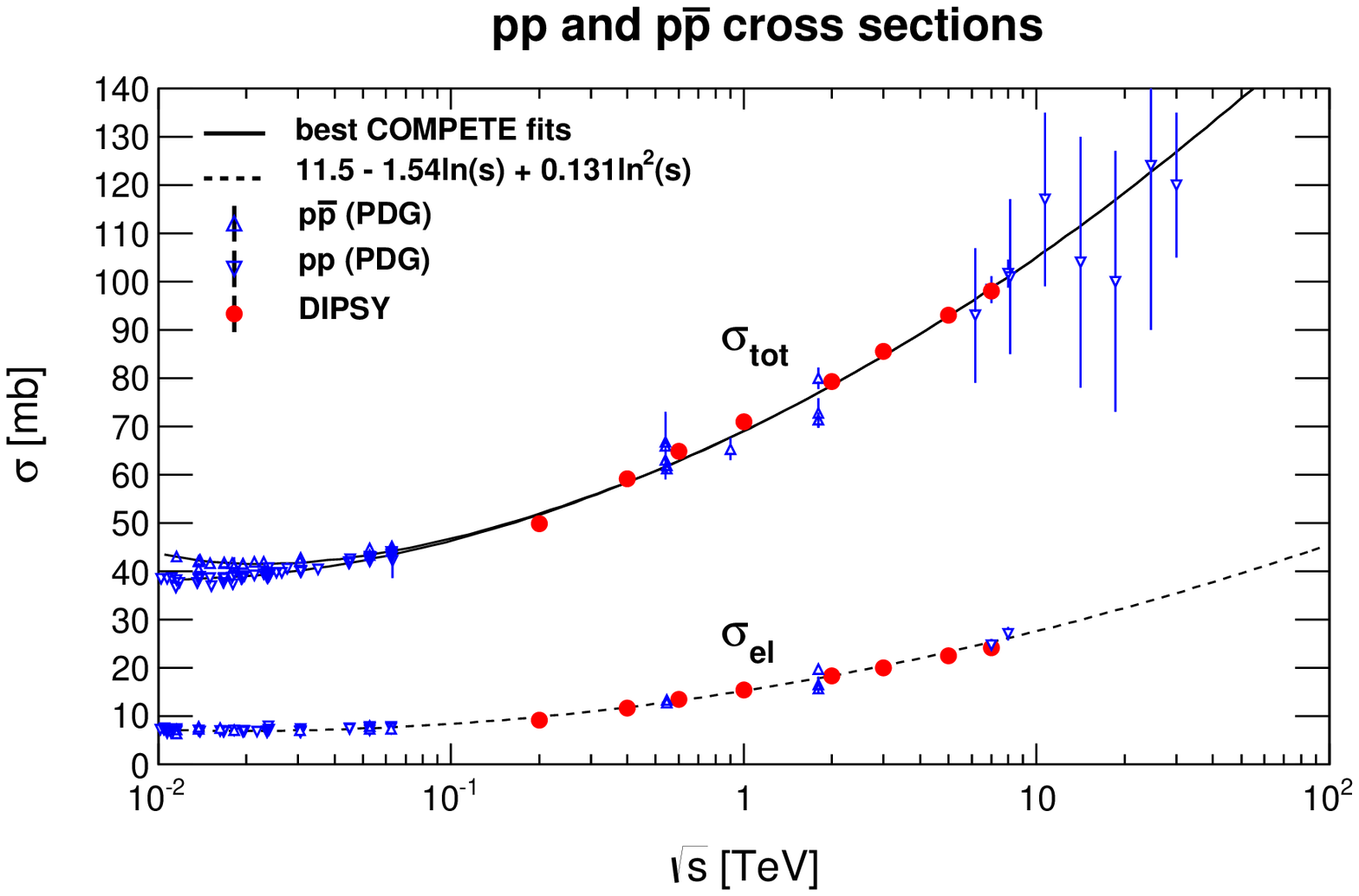,width=0.8\linewidth}
  \caption{\label{fig:pp} Total and elastic cross sections in \pp\ and
    $\mrm{p}\overline{\mrm{p}}$ collisions plotted together with tuned
    \dipsy simulation results (red filled circles).  The total cross
    section data (open triangles) are described by the best fit of the
    COMPETE collaboration~\cite{compete:2002} and the inelastic cross
    sections are also fitted by a phenomenological formula as shown in
    the figure.  }
}

\subsubsection[\pPb\ inelastic cross sections]
{\boldmath \pPb\ inelastic cross sections}

Above the highest RHIC ion collision energy of $\sqrt{s_{NN}} =
200$~GeV, in the region where \dipsy can be reliably applied, there
are only a few heavy ion cross sections measured. One of them is the
\pPb\ total inelastic cross section for which the
CMS~\cite{CMS:2013rta} and LHCb~\cite{LHCb:2012aka} collaborations
have presented preliminary results\footnote{Note that the LHCb value
  required events with at least one track in their acceptance region,
  while the CMS value has been fully corrected to the total inelastic
  cross section. Also ALICE has presented a related result for visible
  cross section \cite{ALICE:2012xs} consistent with LHCb.} as shown in
\figref{fig:pPb-inel}. The \dipsy results, shown by a solid line
connecting the simulated points, agrees well with the data, as does
the two additional theoretical calculations shown from the
EPOS-LHC~\cite{epos-lhc:2006} and QGSJetII-04~\cite{qgsjet-II:2009}
models.

The \dipsy results are mainly driven by the underlying geometrical
description of the nucleus from the GLISSANDO parameterisation, as
discussed in more detail in \sectref{sec:semi-inclusive-cross}.

\FIGURE[t]{
  \centering
  \epsfig{file=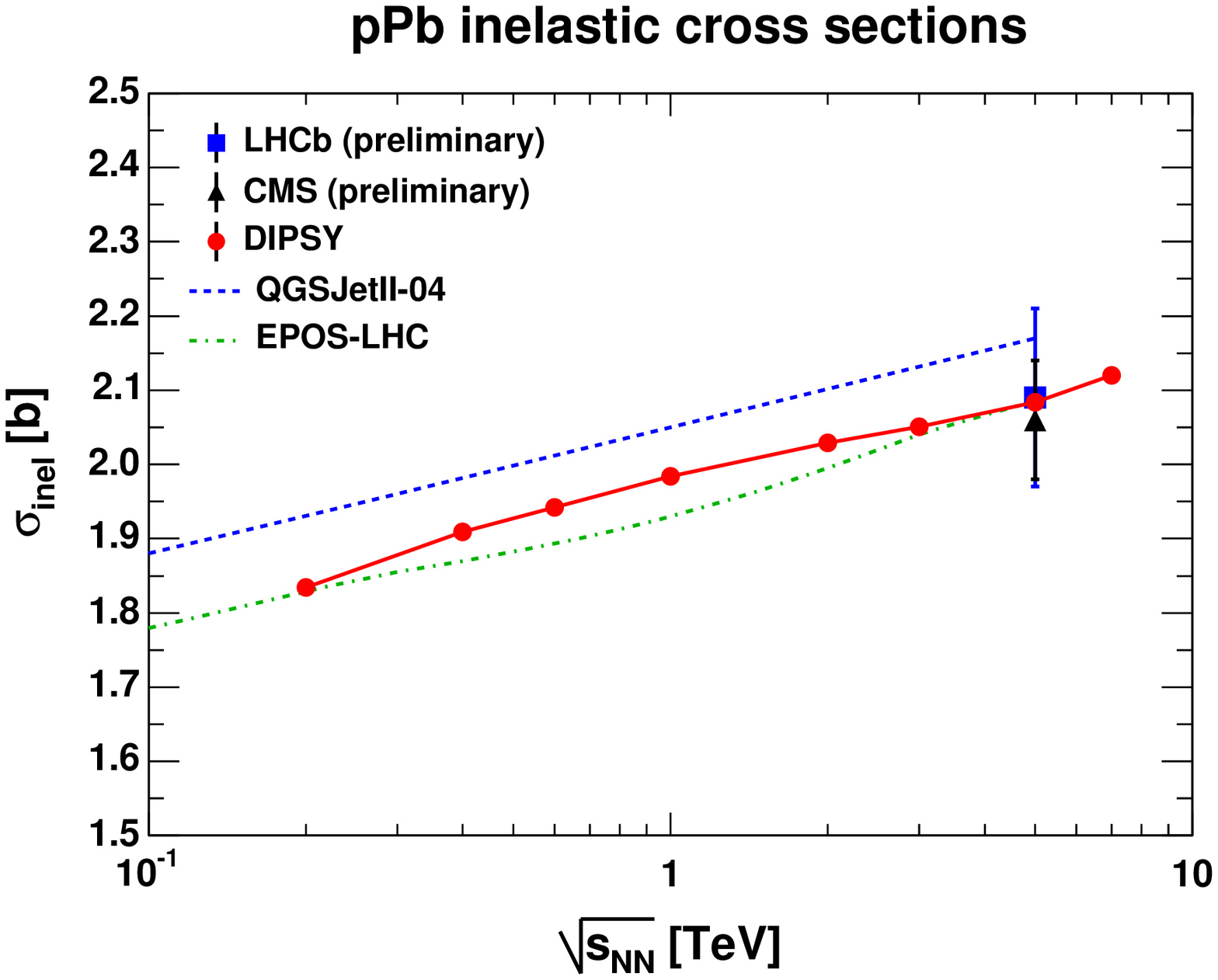,width=0.6\linewidth}

  \caption{\pPb\ inelastic cross sections recently measured by the CMS
    (black triangle) and the LHCb (blue square) collaborations.  The
    solid line guides the eye for the \dipsy results starting from
    RHIC collision energies.  The predictions of
    QGSJetII-04~\cite{qgsjet-II:2009} and
    EPOS-LHC~\cite{epos-lhc:2006} simulations are indicated by dotted
    and dashed lines, respectively.  }
  \label{fig:pPb-inel}
}

\subsection[Predictions for \pA\ and \gA\ total cross sections]
{\boldmath Predictions for \pA\ and \gA\ total cross sections}

\subsubsection[\pA\ total cross sections]{\boldmath \pA\ total cross sections}

In \figref{fig:pA} total cross sections calculated by \dipsy are shown
for collisions of protons with different types of ions. Lead, copper
and oxygen were selected to represent heavy, intermediate, and light
ions to get an overview about the main characteristics of \pA\ cross
sections.  We note that these cross sections grow more slowly with
energy than the \pp\ cross section in \figref{fig:pp}. This is more
clearly seen in \figref{fig:pA-pp}, where we show the ratio of the
total \pA\ cross section to the corresponding total \pp\
cross section. For clarity, we here also normalise this ratio by the
mass number, $A$, for the colliding ion.  If the $NN$ cross section
would be small, this ration should be one. The deviation from one,
which is larger for the heavy lead nucleus and for higher energy, is
thus a related to the average number of $NN$ collisions in a
single \pA\ event.

\FIGURE[t]{
  \centering
  \epsfig{file=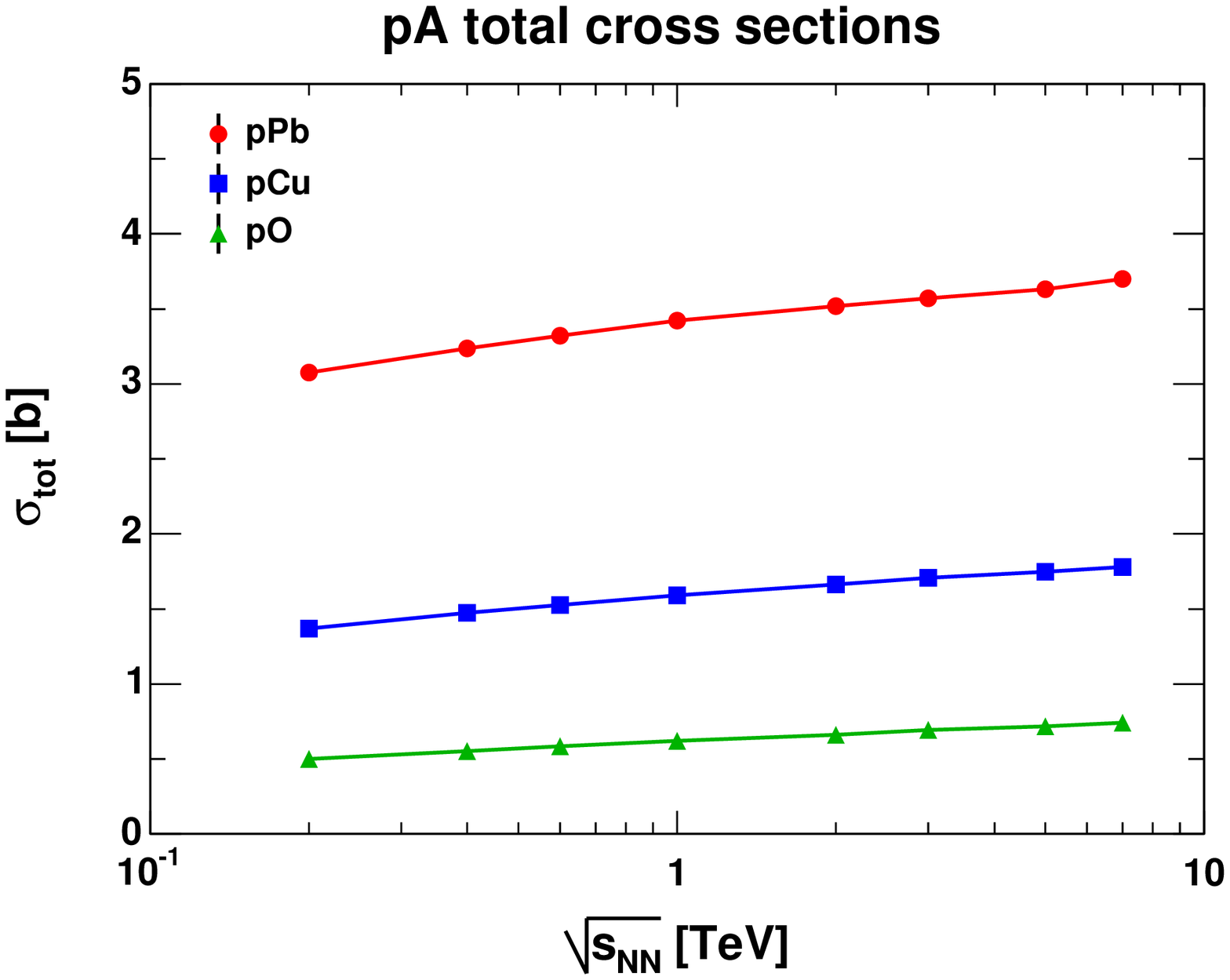,width=0.6\linewidth}
  \caption{Total cross sections for \pA\ collisions obtained from the
    \dipsy Monte Carlo model calculations.  The results are shown as a
    function of the nucleon--nucleon centre-of-mass energy, covering a
    range from the upper RHIC energies to the available LHC energies.}
  \label{fig:pA}
}

For a completely black nucleus, the cross section is expected to scale
with the area overlap for the two projectiles, \ie\ $\propto
(R_A+R_p)^2$. Approximating $R_A\sim A^{1/3}R_p$ will thus lead to a
scaling $\sim R_p^2(A^{1/3}+1)^2$. Figure \ref{fig:pA} also shows the
\pA\ total cross sections scaled by $(A^{1/3}+1)^2$.  We can here see
that, as expected, this area scaling works better for heavy nuclei and
high energy, when the black limit is approached.  Analogously the
contribution from volume effects $\propto A$ are more visible for
smaller nuclei and lower energy.

\FIGURE[t]{  
  \centering
  \epsfig{file=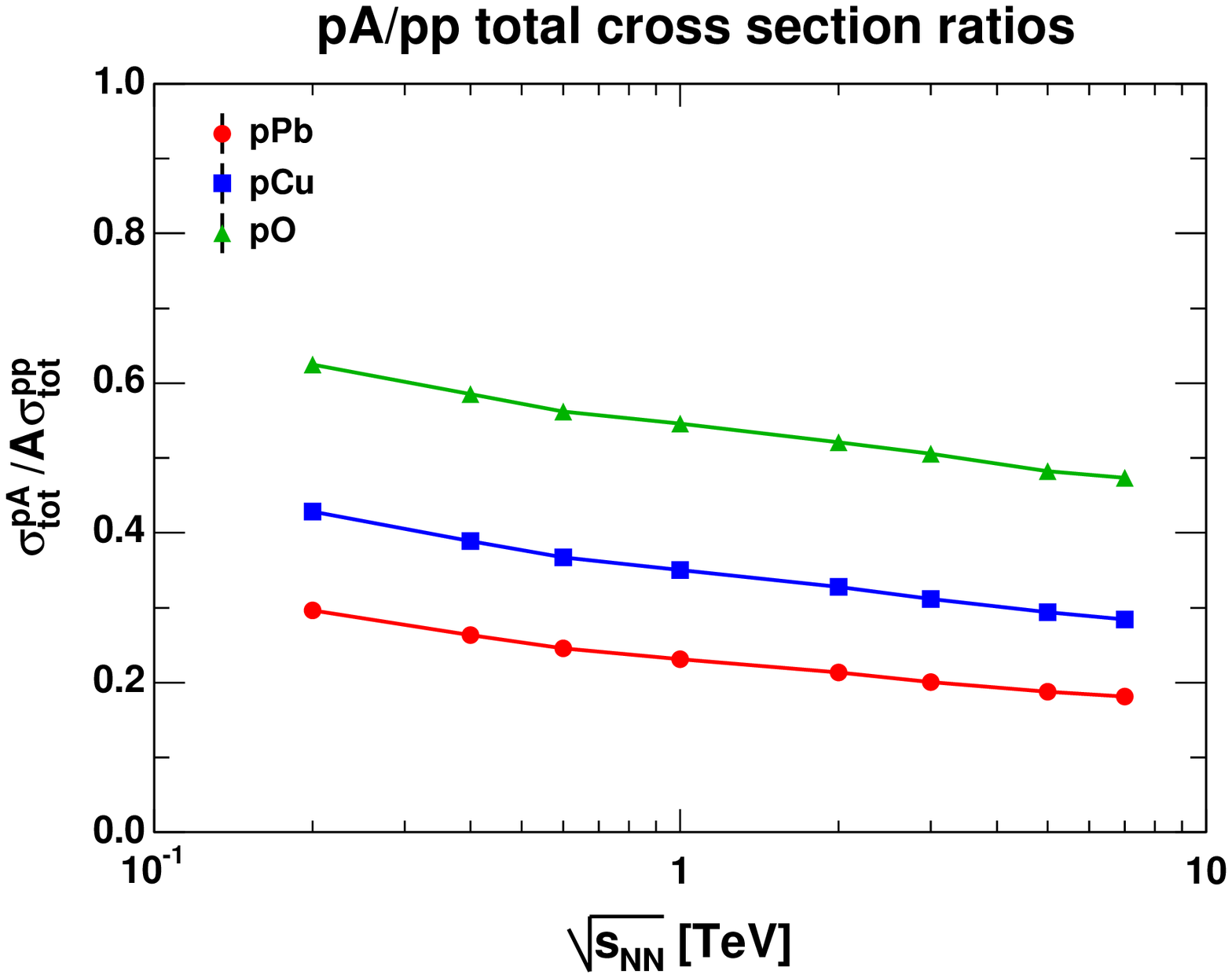,width=0.45\linewidth}%
  \epsfig{file=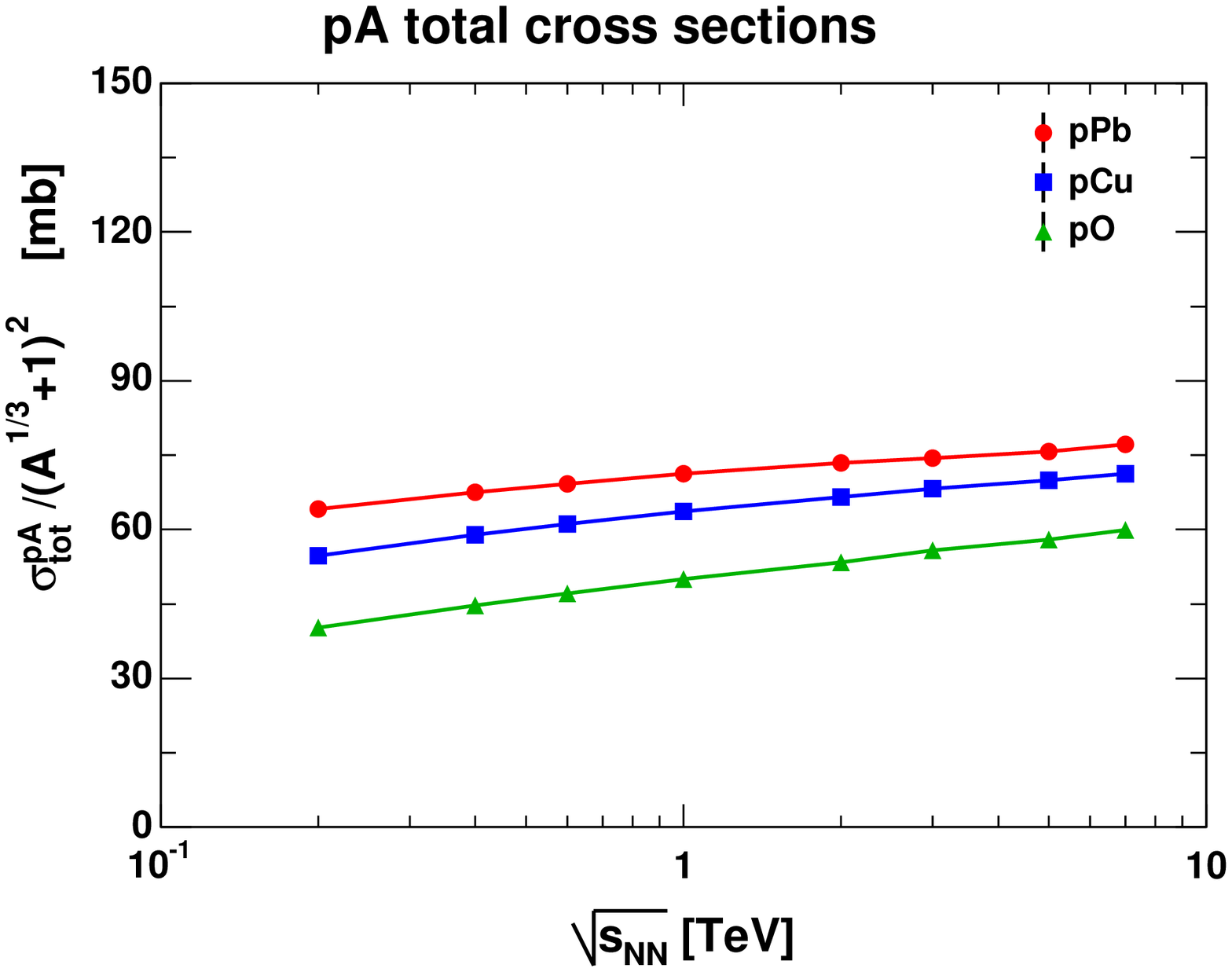,width=0.45\linewidth}
  \caption{\dipsy predictions for the total \pA\ cross sections scaled
    by $A\sigma\subtot^{pp}$ (left), and by $(A^{1/3}+1)^2$
    (right). For the mass-number scaling the ratio for \pO\ collisions
    closer to 1 than for \pPb\ collisions, indicating, that volume
    effects are more important for lighter nuclei and at lower
    colliding energies, as far as the non-linear nuclear effects
    implemented in \dipsy are considered, while the opposite is found
    for the surface scaling.}
  \label{fig:pA-pp}
}

\subsubsection[\gA\ total cross sections]{\boldmath \gA\ total cross sections}

When a (virtual) photon fluctuates into a $q\bar{q}$ pair, it forms a
colour dipole, which interacts with the nucleus in the same way as a
dipole in a colliding proton.  Thus it produces a dipole cascade,
which is comparatively long-lived, and frozen during the interaction
with the nucleus. In \figrefs{fig:eO-tot}-\fig{fig:eAu-tot} we show
the total cross section for \gA\ collisions in the energy range
$\sqrt{s_{\gA}}$ between 0.1 and 7 TeV, similar to the energy range
studied above for \pA\ collisions. Results are presented for photon
virtualities $Q^2$ = 2.5, 5, 25 and 100~GeV$^2$ colliding with O, Cu,
and Au nuclei. We also present the results after scaling with the
\gp\ cross section and the mass number $A$.
We have used the parameters tuned to the \pp\ cross sections in
\sectref{sec:pp-tune}, and we have checked that we reproduce data on
\gp\ cross sections to a reasonable degree. This means, however, that
the \gA\ cross sections presented here should not be thought of as
true precision predictions, and we will here only discuss the
qualitative behavior of the cross sections and their ratios.

\FIGURE[t]{  
  \centering
  \epsfig{file=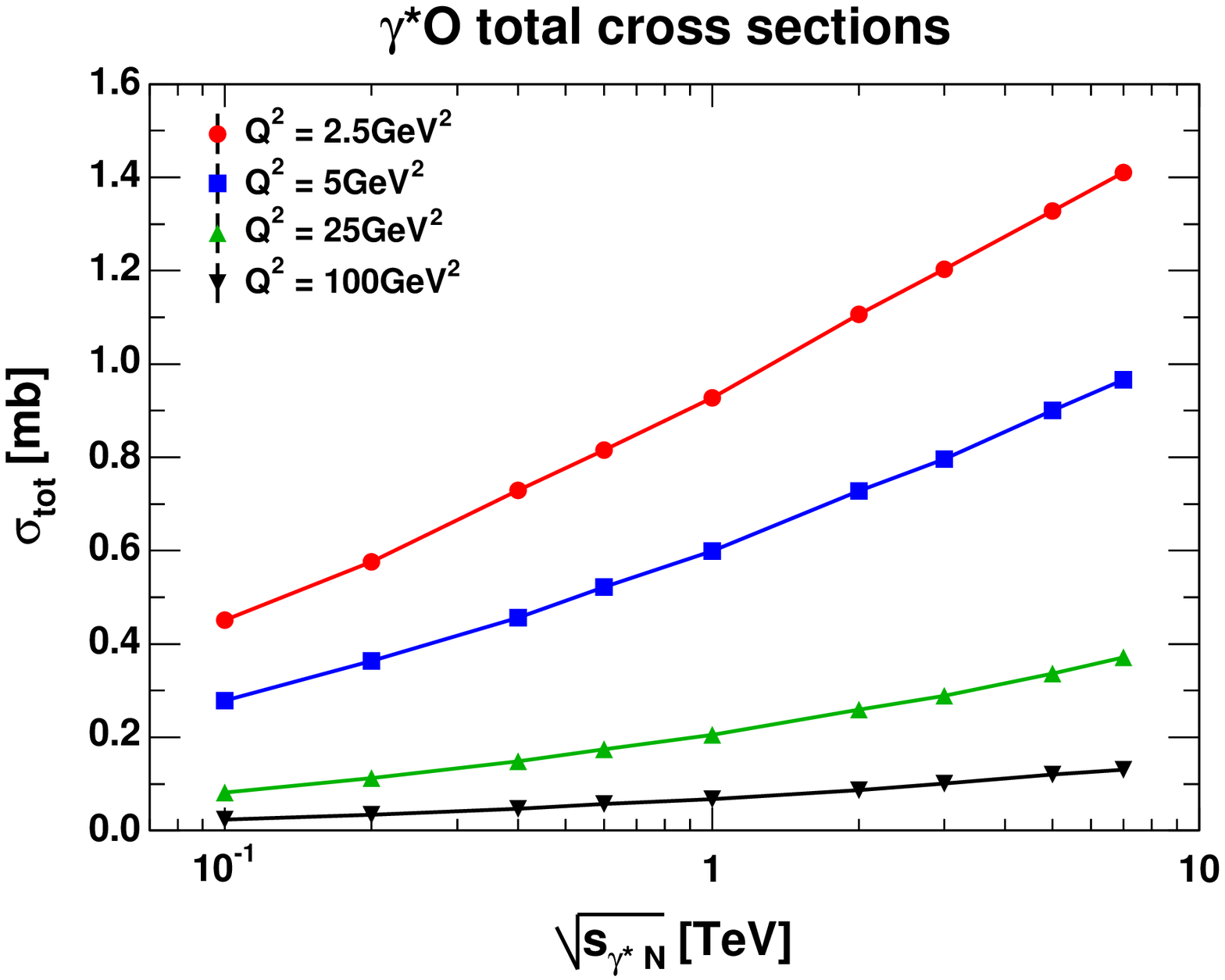,width=0.48\linewidth}
  \epsfig{file=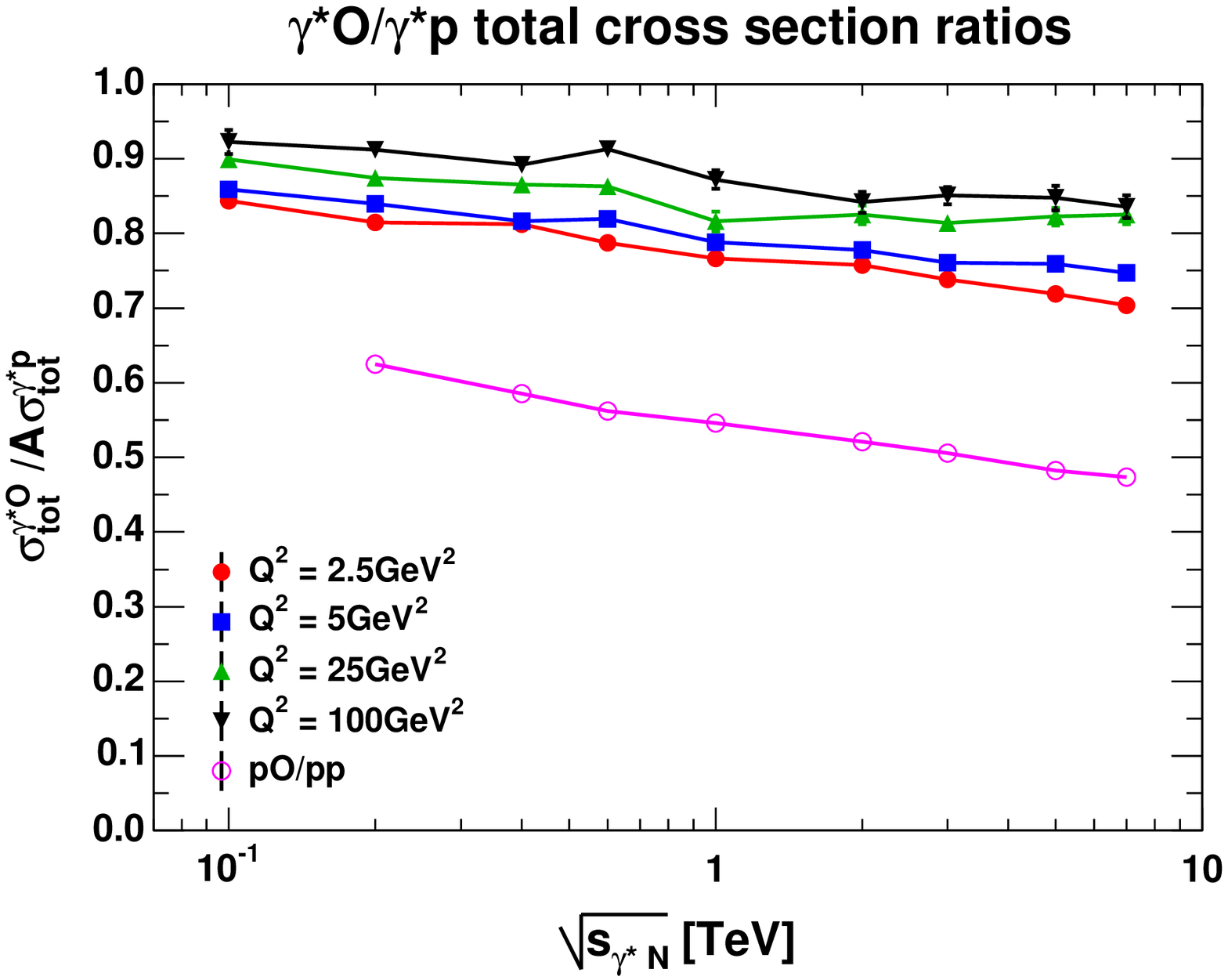,width=0.48\linewidth}
  \caption{ Left panel indicates \dipsy predictions for the total
    cross sections of \gO\ reactions as a function of the
    photon virtuality $Q^2$ and the available energy
    $\sqrt{s_{\gN}}$.  The right panel compares these
    calculations to \gp\ reactions, scaled also by the
    mass number $A$.  For comparison we also show the ratio \pO/\pp.  }
  \label{fig:eO-tot}
}

We note that the dipole--nucleon cross section is suppressed compared
to the $NN$ interaction, as an effect of colour
transparency. Therefore the cross sections scale more closely with the
volume $\sim A$ than with the area, as was the case for \pA\
collisions.  For the same reason it also grows with energy,
approximately proportional to the \gp\ cross section. As seen in the
figures, this scaling is most clear for larger $Q^2$ and smaller $A$,
while saturation effects become more noticeable for lower $Q^2$ and
heavier nuclei.

The \gA\ cross section is also suppressed by the factor $\alpha_{EM}$
in the photon--quark coupling. Note that the cross sections are here
measured in millibarn, instead of barn, which was used above for the
\pA\ cross sections. However, as the dipole state is frozen during the
collision, this extra suppression has no effect on the scaling
behaviour discussed here.

\FIGURE[t]{  
  \centering
  \epsfig{file=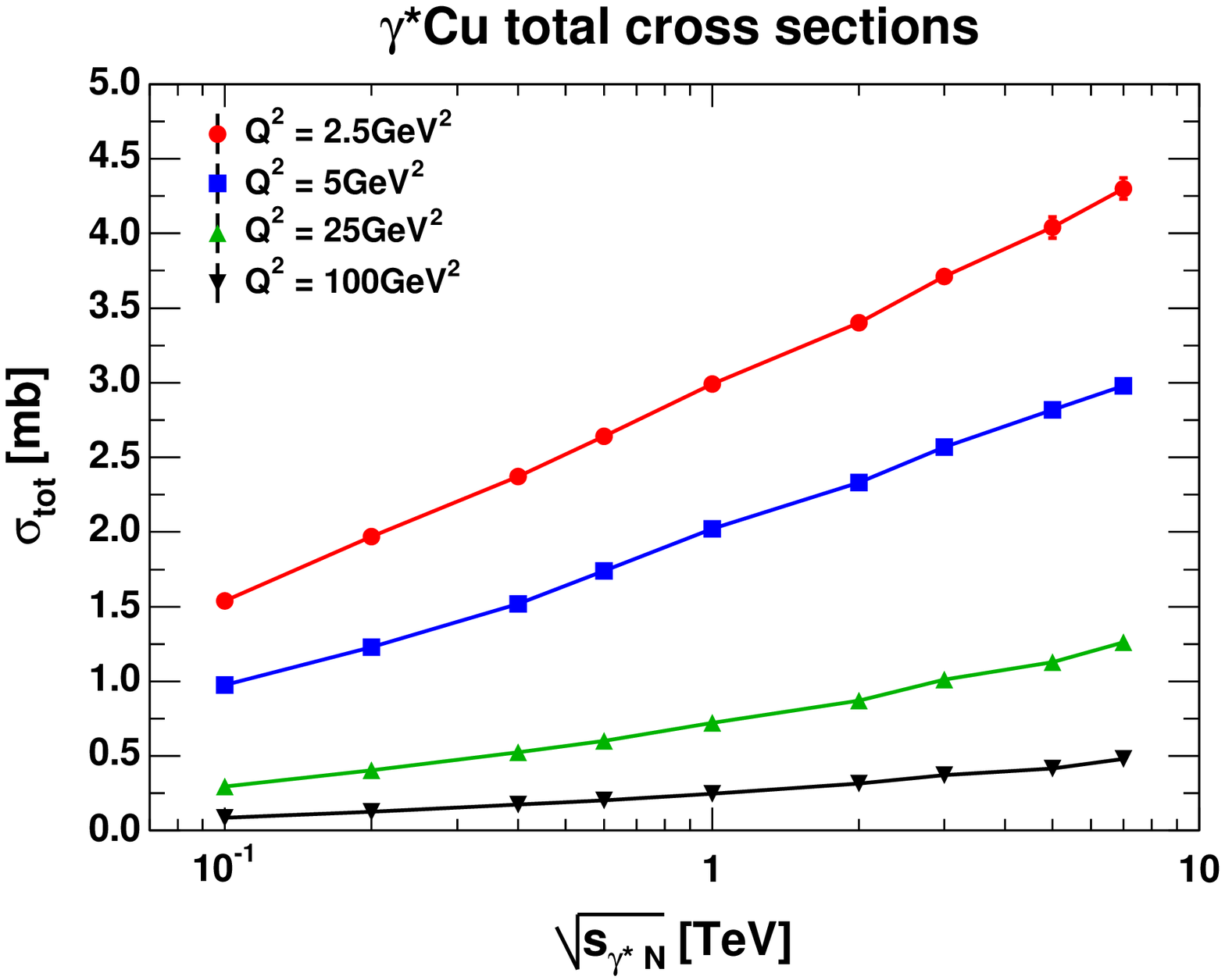,width=0.48\linewidth}
  \epsfig{file=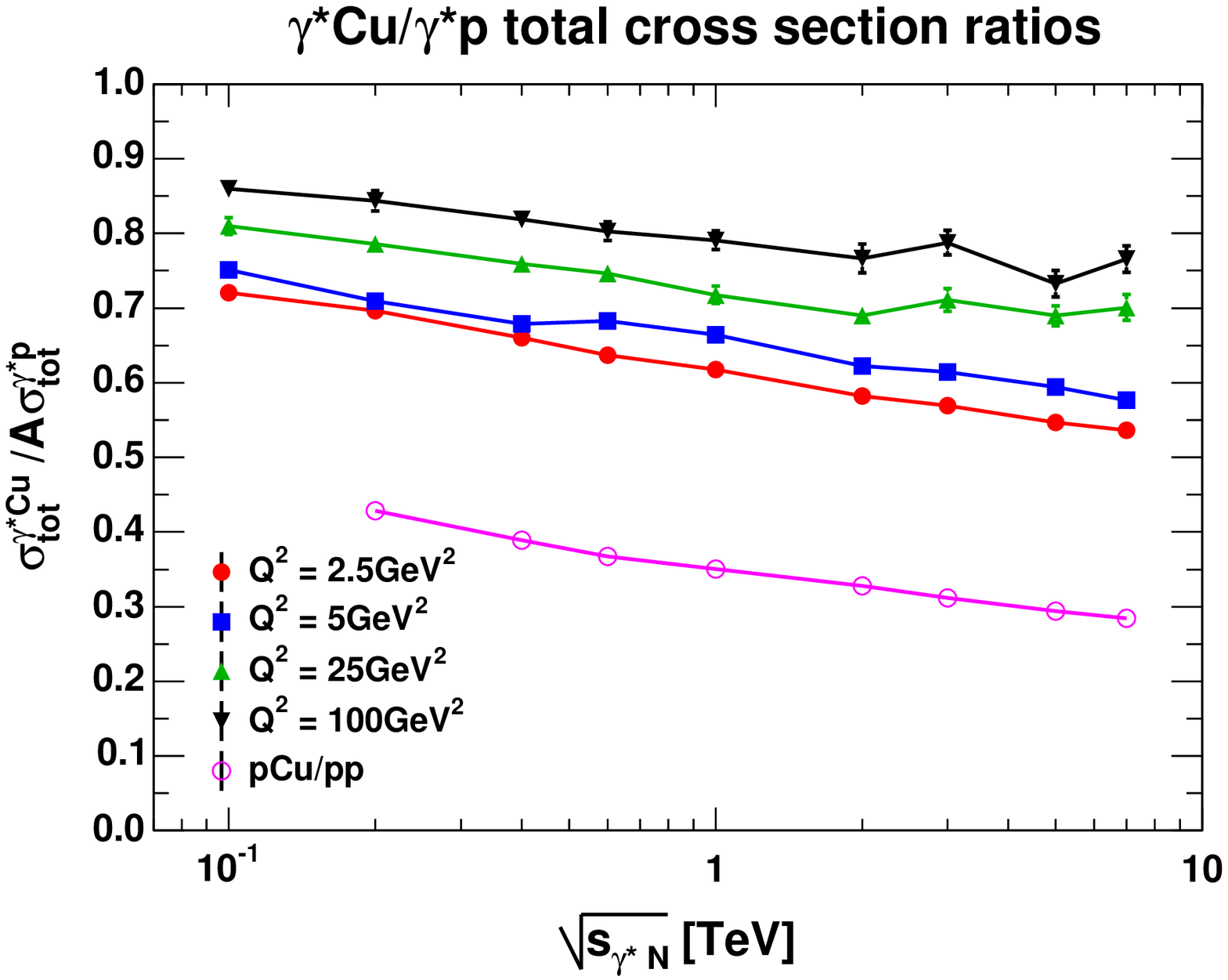,width=0.48\linewidth}
  \caption{
    Same as \figref{fig:eO-tot}  but for Cu nuclei.
  }
  \label{fig:eCu-tot}
}

\Figref{fig:eCu-tot} indicates the results of a \dipsy simulation for
the intermediately heavy Cu nuclei that can be compared in a
straightforward way with the results shown on \figref{fig:eO-tot}.
From this comparison one can observe, that $\sigma\subtot^{\gCu} / A
\sigma\subtot^{\gp}< \sigma\subtot^{\gO} / A \sigma\subtot^{\gp}$ .
This suggests that even for intermediately heavy nuclei such as Cu, it
is more difficult for the dipoles to find a hole and to pass through
these nuclei, as compared to lighter nuclei such as O.

\FIGURE[t]{
  \centering
  \epsfig{file=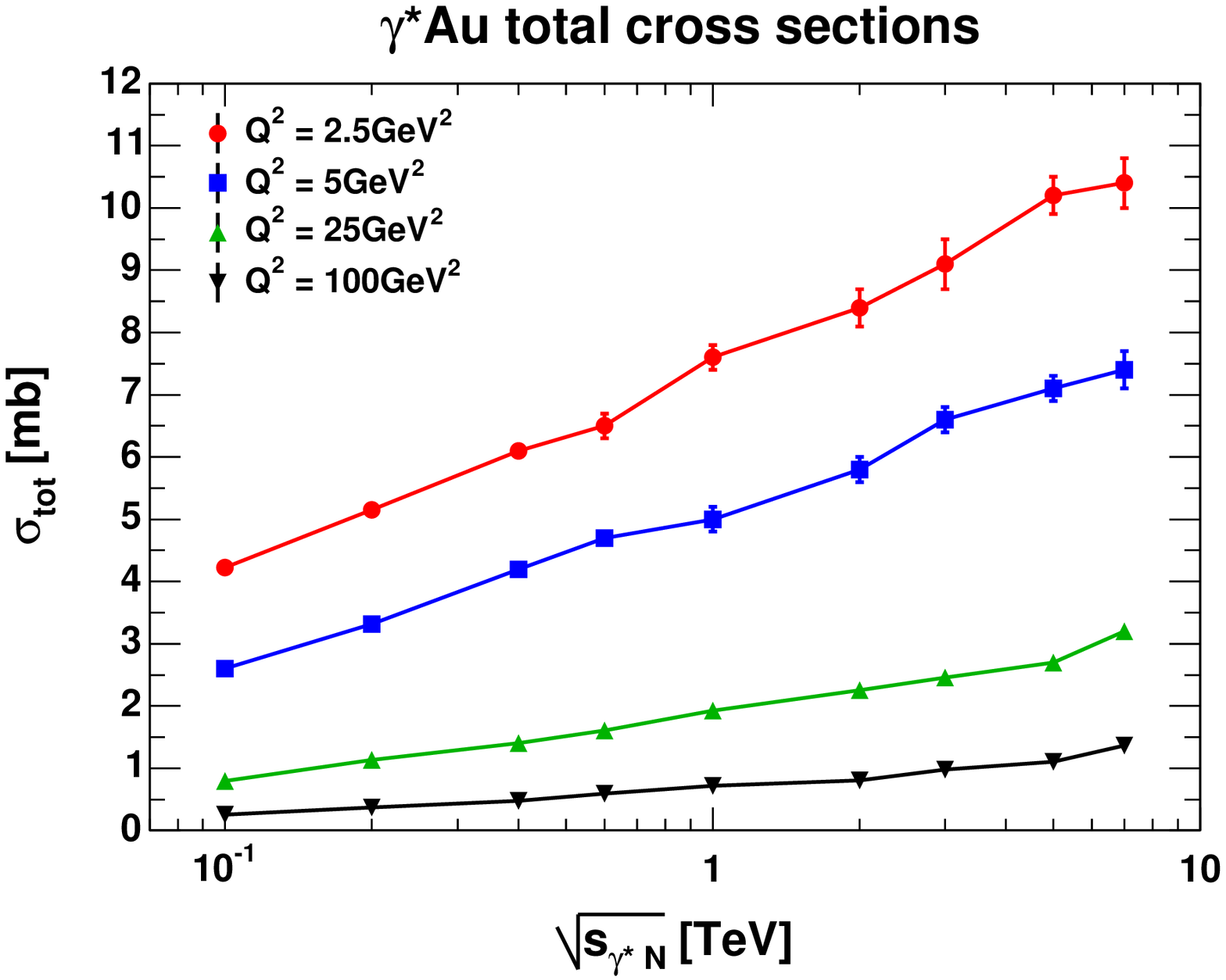,width=0.48\linewidth}
  \epsfig{file=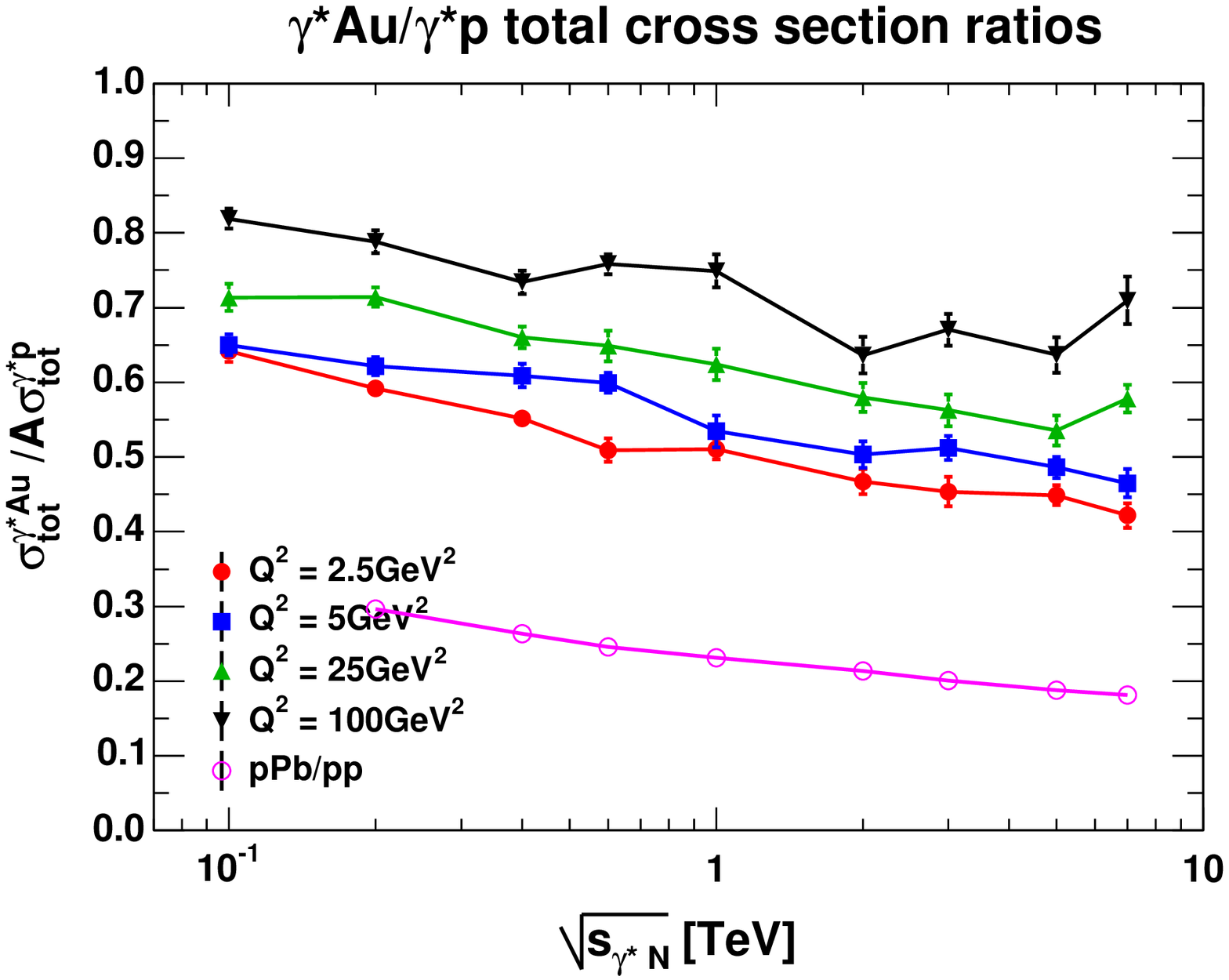,width=0.48\linewidth}
  \caption{
    Same as \figref{fig:eO-tot}  but for Au nuclei.
  }
  \label{fig:eAu-tot}
}

\Figref{fig:eAu-tot} indicates that in \gAu\ reactions the mass number or volume scaling is even worse approximation than in \gCu\ reactions.
This indicates that for large nuclei and for large $\sqrt{s}$ the nuclei  
become asymptotically black even for a small dipole coming from a 
virtual photon, $\gamma^\star$. Nevertheless, when compared to calculations
with similar centre-of-mass energies for \pPb\ collisions, one can observe
that  virtual photons can pass through these large nuclei \Au\ or \Pb\ 
more easily as compared to the collision of protons with the same target.

\subsection[Semi-inclusive \pA\ cross sections and comparisons with Glauber models]{\boldmath Semi-inclusive \pA\ cross sections and comparisons with Glauber models}
\label{sec:semi-inclusive-cross}

Besides colour coherence effects, to be discussed in section
\ref{sec:sat-eff}, the most important feature of the \dipsy model for
nucleus collisions is its treatment of correlations, and fluctuations,
not only between the nucleons, but also inside the nucleons and
between partons in different nucleons. To investigate the observable
consequences of this we will in this section compare some results for
semi-inclusive cross sections in \dipsy with those of standard Glauber
calculations, where mainly fluctuations in the positions of nucleons
and short range nuclear correlations between nucleons are considered.

In \figref{fig:pA-glau-ND} we present \dipsy predictions for the
inelastic, non-diffractive cross sections for \pA\ collisions as a
function of collision energy, comparing to a Glauber calculation using
the black disc with the same nucleon distribution (see
\sectref{sec:nucleongliss}) as input. Clearly the results are quite
similar with differences only at the percent level. However, if we
look at total \pA\ cross sections in \figref{fig:pA-glau-tot}, the
differences are much larger, almost up to 30\% for small nuclei and
lower energies. The differences here are mainly in the way how
fluctuations are treated.

\FIGURE[t]{
  \centering
  \epsfig{file=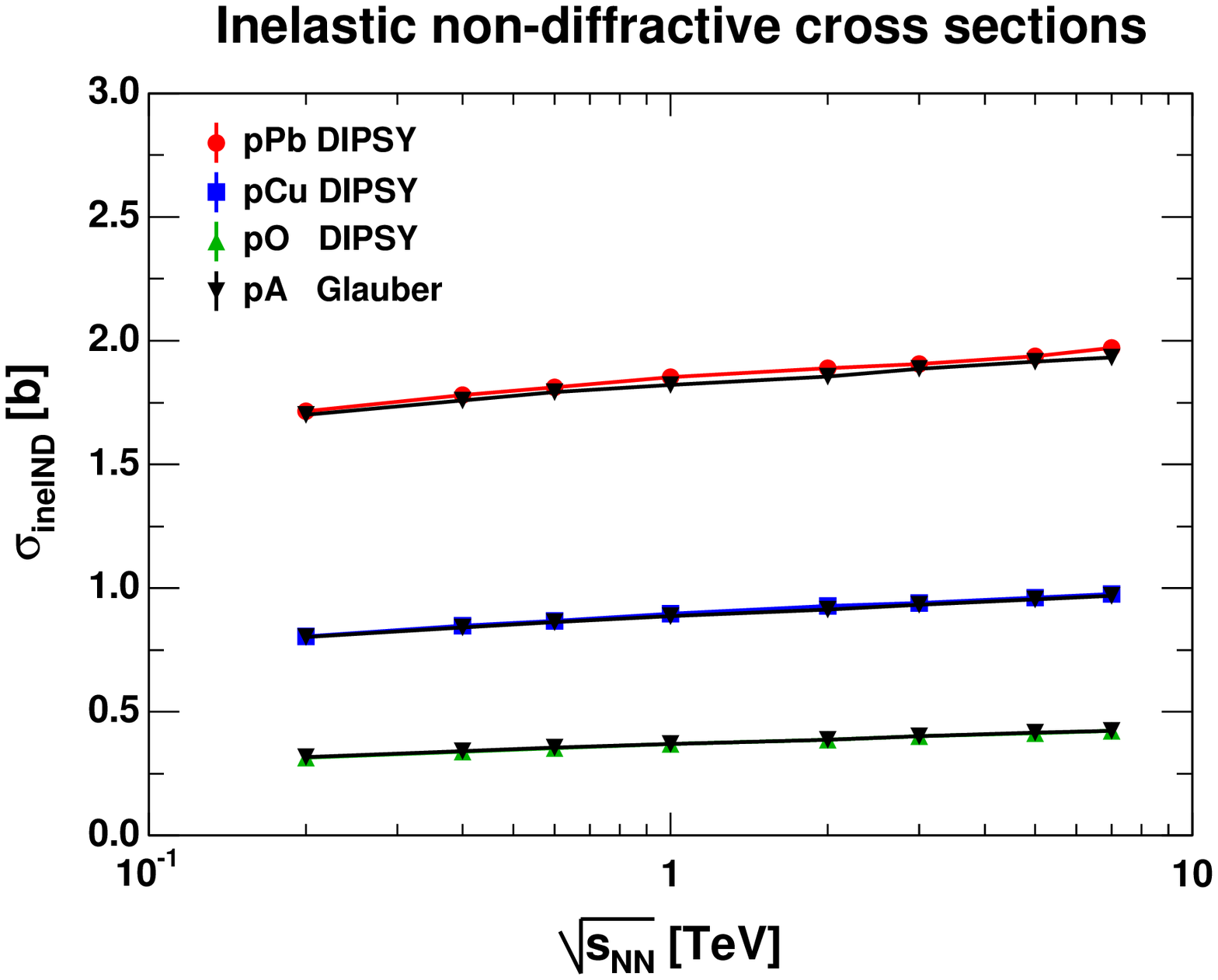,width=0.45\linewidth}
  \epsfig{file=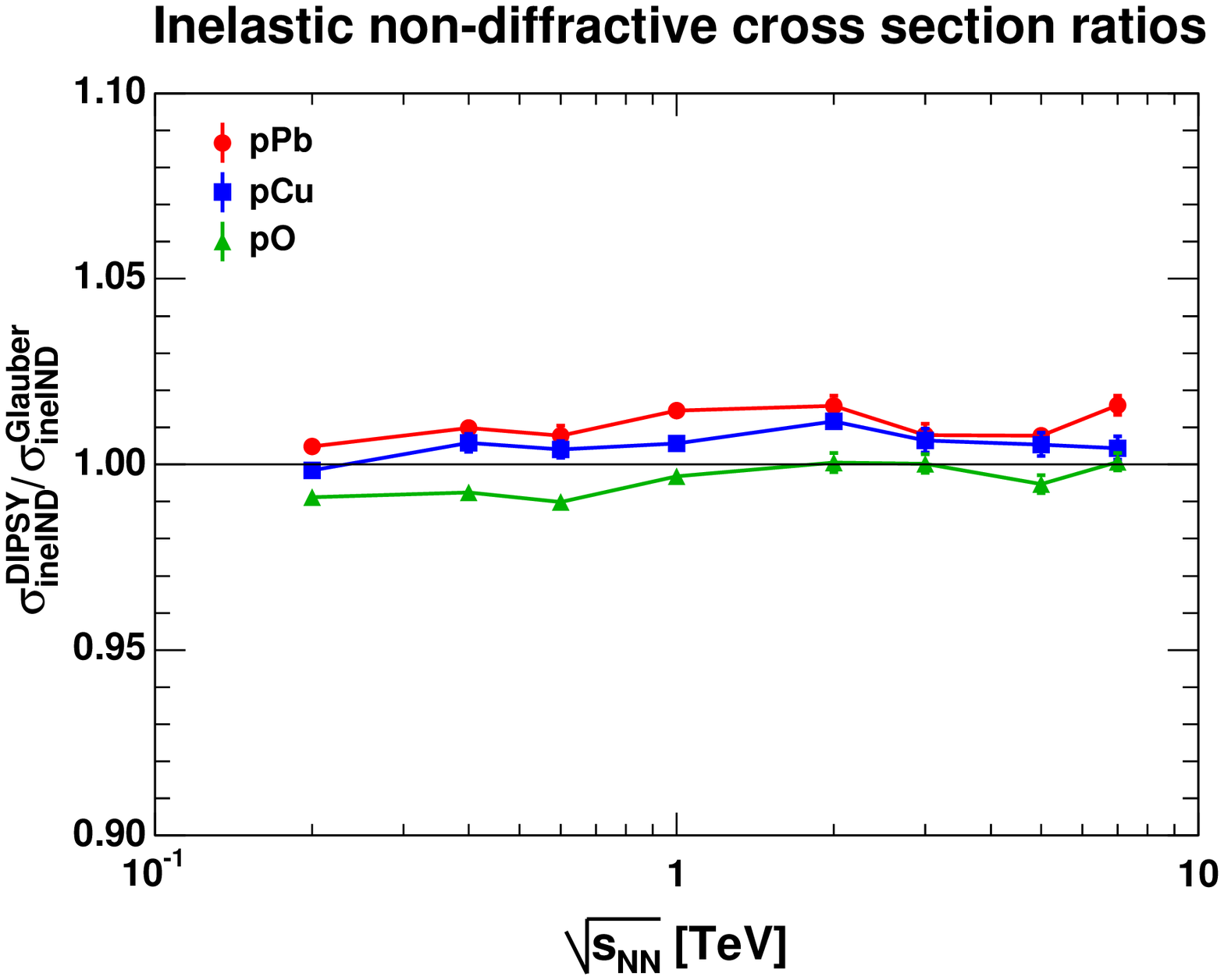,width=0.45\linewidth}
  \caption{On the left panel Glauber Monte-Carlo and \dipsy
    calculations are compared for the $\sigma\subinelnd^{\pA}$
    inelastic non-diffractive cross sections.  In these calculations,
    the \dipsy results for $\sigma\subinelnd$ \pp\ cross sections were
    used, as input, in the Glauber calculations. The Glauber Model MC
    results were obtained in the black-disc approach.  The right panel
    shows the ratios of the \dipsy calculation to the Glauber Monte
    Carlo, indicating that the effects of correlations and
    fluctuations contribute to these variables on the typical 1\%
    scale.  }
  \label{fig:pA-glau-ND}
}

\FIGURE[t]{
  \centering
  \epsfig{file=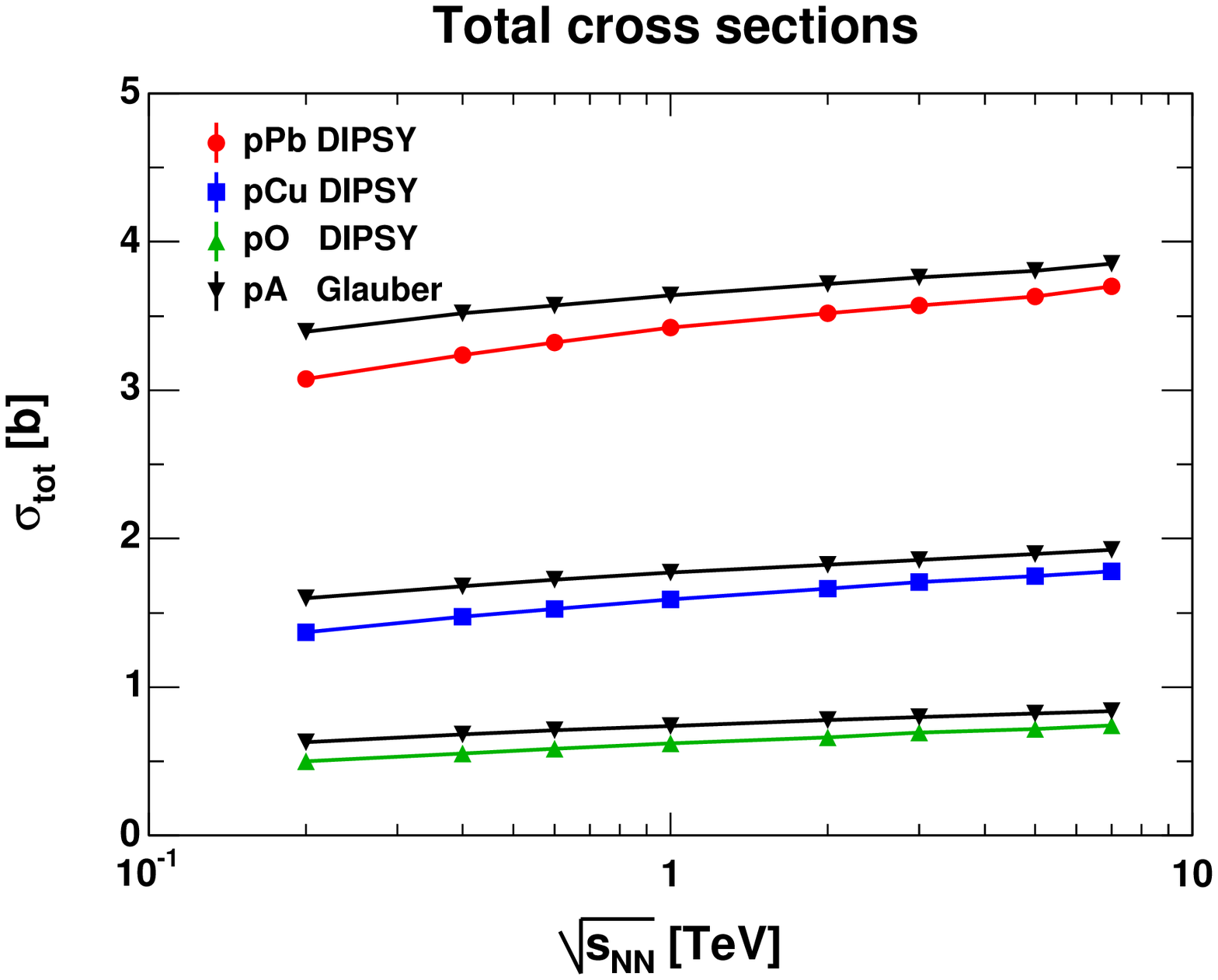,width=0.45\linewidth}
  \epsfig{file=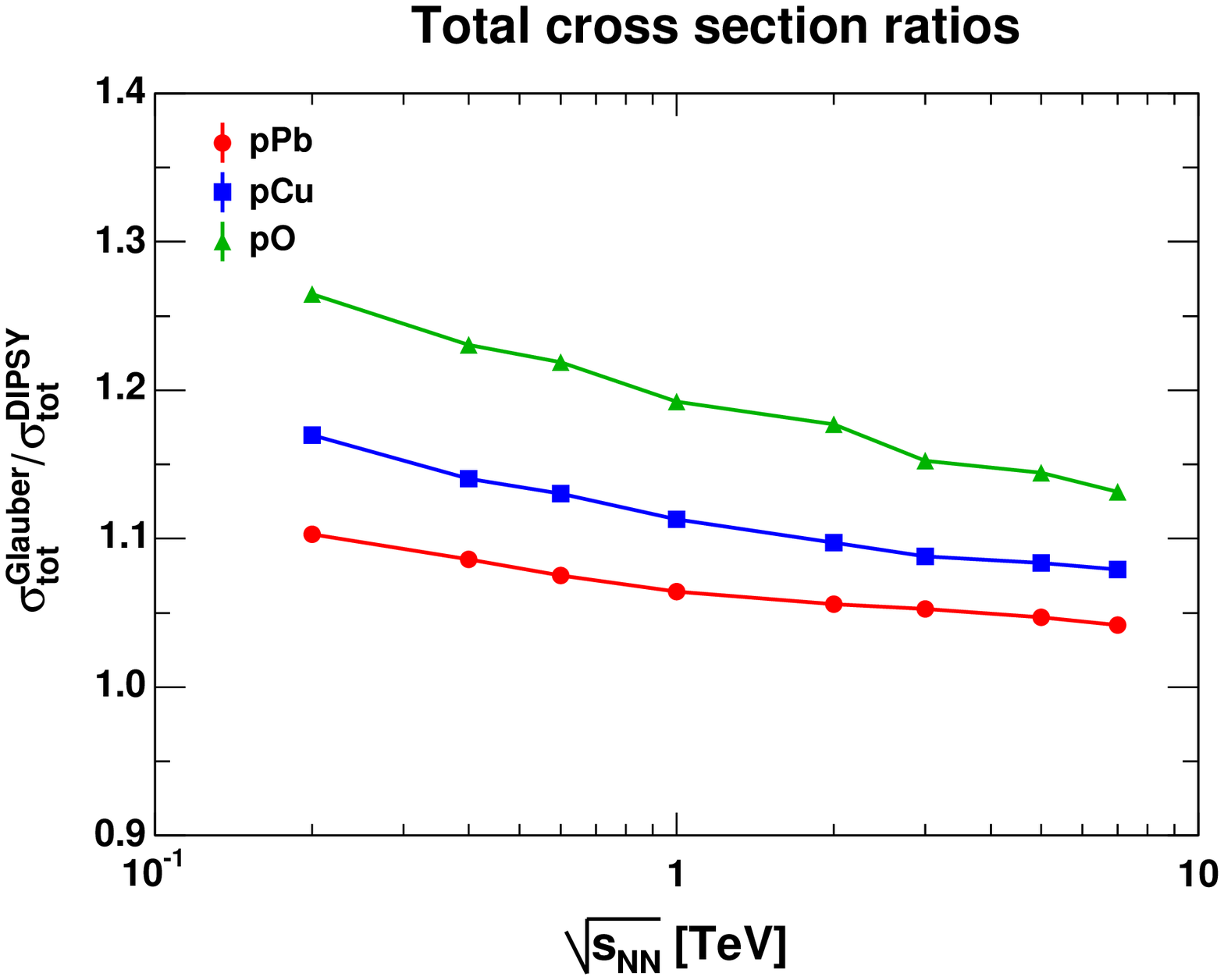,width=0.45\linewidth}
  \caption{Left panel indicates a comparison of $\sigma\subtot^{\pA}$
    total cross section calculations from \dipsy to the corresponding
    Glauber Model MC results, using the black-disc approach. For
    Glauber, the same $\sigma\subinelnd$ \pp\ cross sections were
    used, as input, that \dipsy calculated.  Right panel indicates the
    ratios of $\sigma\subtot^{\pA}$ total cross sections calculated by
    \dipsy to the Glauber Model MC in black-disc approach, indicating
    that although the inelastic cross sections are within 1\% the same
    in the two approach, the predictions for the total cross section
    differ by a rather significant value that is about $20-30$\% at
    RHIC energies and decreases to $10-15$\% at LHC energies.  }
  \label{fig:pA-glau-tot}
}

It should be noted that the input radius to the Glauber calculation is
here taken from the inelastic, non-diffractive \pp\ cross sections, \cf\
\eqref{eq:blackdisc}, using the value obtained from \dipsy. However,
looking at \eqref{eq:crossectionsGl}, it is clear that within the
black disc approximation we could also have taken the total or elastic
\pp\ cross sections to define the input radius. In fact, if we had taken
$\sigma\sup{pp}\sub{tot}=2\pi R^2$ to define the input radius, the
difference w.r.t.\ \dipsy would be much smaller.  On the other hand
the differences for the inelastic, non-diffractive cross sections
would then be larger.

As pointed out in \sectref{sec:diff}, the observables most sensitive
to fluctuations are related to diffraction. For the black-disc Glauber
calculations in \eqref{eq:crossectionsGl}, the \pp\ diffractive cross
section is zero. For \pA\ collisions, however, the Glauber calculation
will result in a diffractive component, which is then entirely related
to the fluctuations in the nucleon distribution within the
nucleus. This can also be thought of as the elastic scattering of the
proton projectile with one or more of the nucleons, resulting in a
diffractive excitation of the nucleus. It is therefore interesting to
compare this to \dipsy, where more fluctuations are taken into
account.

\newcolumntype{d}{D{.}{.}{3}}
\setlength{\tabcolsep}{5pt}
\def\na{\multicolumn{1}{c|}{-}}
\TABLE[t]{
  \centering
  \begin{tabular}{|l|d|d|d|d|d|d|d|d|d|d|}
    \hline
    Model & \multicolumn{2}{c|}{\dipsy}  & \multicolumn{2}{c|}{Black disc} &
    \multicolumn{2}{c|}{Black disc} & \multicolumn{2}{c|}{Black disc} &
    \multicolumn{2}{c|}{Grey disc}\\
    & \multicolumn{2}{c|}{}  & \multicolumn{2}{c|}{($\sigma\subtot$)} & 
    \multicolumn{2}{c|}{($\sigma\subineltot$)} &
    \multicolumn{2}{c|}{($\sigma\subinelnd$)} &
    \multicolumn{2}{c|}{($\sigma\subtot,\sigma\subel$)}\\
    $\sqrt{s_{\NN}}$ {(TeV)} & \multicolumn{1}{c|}{5} &
    \multicolumn{1}{c|}{10} & \multicolumn{1}{c|}{5} &
    \multicolumn{1}{c|}{10} & \multicolumn{1}{c|}{5} &
    \multicolumn{1}{c|}{10} & \multicolumn{1}{c|}{5} &
    \multicolumn{1}{c|}{10} & \multicolumn{1}{c|}{5} &
    \multicolumn{1}{c|}{10} \\
    \hline
    $\sigma\subtot$\hfill (b) &
    3.54 & 3.62 & 3.50 & 3.58 & 3.88 & 3.95 & 3.73 & 3.80 & 3.69 & 3.77 \\
    $\sigma\subineltot$\hfill (b) & 
    2.04 & 2.07 & 1.95 & 1.98 & 2.14 & 2.17 & 2.06 & 2.09 & 2.07 & 2.11 \\
    $\sigma\subinelnd$\hfill (b) & 
    1.89 & 1.92 & 1.75 & 1.79 & 1.94 & 1.98 & 1.86 & 1.90 & 1.84 & 1.89 \\
    $\sigma\subel$\hfill (b) & 
    1.51 & 1.55 & 1.55 & 1.60 & 1.73 & 1.78 & 1.66 & 1.70 & 1.62 & 1.66 \\
    $\sigma\subsda$\hfill (b) & 
    0.085 & 0.086  & 0.198 & 0.192 & 0.204 & 0.198 & 0.200 & 0.195 & 0.083 & 0.085 \\
    $\sigma\subsdp$\hfill (b) & 
    0.023 & 0.024  & \na & \na & \na & \na & \na & \na & \na & \na \\
    $\sigma\subdd$\hfill (b) & 
    0.038 & 0.038  & \na & \na & \na & \na & \na & \na & 0.142 & 0.137 \\
    $\sigma\subqel$\hfill (b) & 
    1.59 & 1.64 & 1.75 & 1.79 & 1.94 & 1.98 & 1.86 & 1.90 & 1.70 & 1.75 \\
    $\sigma\subqel/\sigma\subineltot$ & 
    0.78 & 0.79 & 0.90 & 0.90 & 0.91 & 0.91 & 0.90 & 0.91 & 0.82 & 0.83 \\
    $\sigma\subinelnd/\sigma\subtot$ & 
    0.53 & 0.53 & 0.50 & 0.50 & 0.50 & 0.50 & 0.50 & 0.50 & 0.50 & 0.50 \\
    \hline
  \end{tabular}
  \caption{\dipsy predictions for different (semi-) inclusive \pPb\
    cross sections (given in units of barns) at LHC energies
    ($\sqrt{s^{NN}}=$~5 and 10~TeV) compared to different Glauber
    calculations. The input radius to the black disc calculations were
    given by $\pi R^2=\sigma\sup{pp}\subtot/2$,
    $\sigma\sup{pp}\subinelnd$, and $\sigma\sup{pp}\subineltot$. For
    the Grey disc both $\sigma\sup{pp}\subtot$ and
    $\sigma\sup{pp}\subel$ were used (see \eqref{eq:greydisc}).  In
    all cases using these \pp\ cross sections obtained from \dipsy.
    The cross sections shown are the total ($\sigma\subtot$), the
    inelastic including diffractive excitation ($\sigma\subineltot$),
    the inelastic non-diffractive ($\sigma\subinelnd$), the elastic
    ($\sigma\subel$), the single diffractive excitation of the nucleus
    ($\sigma\subsda$) and proton ($\sigma\subsdp$), the double
    diffractive excitation ($\sigma\subdd$), and the quasi-elastic
    scattering ($\sigma\subqel$ in \eqref{eq:qel}). Also shown are the
    cross section ratios, $\sigma\subqel/\sigma\subineltot$ and
    $\sigma\subinelnd/\sigma\subtot$.  The statistical uncertainty in
    the calculations are below a percent, except for the diffractive
    excitation cross sections, where the uncertainty is around 3\%.}
  \label{tab:semiincxsec}
}

In table~\ref{tab:semiincxsec} we present predictions from \dipsy for
different (semi-) inclusive \pPb\ cross sections at LHC energies
($\sqrt{s^{NN}}\approx5$ and 10~TeV) compared to different Glauber
calculations. The input for the black disc radii were taken from the
\pp\ cross sections calculated by \dipsy, and for all cross sections we
find a trivial increase as the radius becomes larger
($\sigma\sup{pp}\subtot/2<\sigma\sup{pp}\subinelnd<\sigma\sup{pp}\subineltot$).
As noted before the inelastic, non-diffractive cross section is close
to \dipsy, when using the \pp\ inelastic, non-diffractive cross section
for the black-disc radius. We note, however, that the predicted cross
sections for the diffractive dissociation of the nucleus are quite
insensitive to the input radius, and are for a black disc in all three
cases much larger than the predictions from \dipsy. This effect also
shows up in the quasi-elastic cross sections
(\cf~\sectref{sec:diff}). The reason for this can be found in the fact
that all black-disc calculations over-estimate the elastic
nucleon--nucleon cross section, resulting in a larger probability to
quasi-elastically excite the nucleus.

It is therefore interesting to also compare with a grey-disc Glauber
calculation (\cf~\eqref{eq:greydisc}), where both the elastic and the
total \NN\ cross sections are set to the same as in \dipsy. In
table~\ref{tab:semiincxsec} we see that the nucleus dissociation cross
section in this case is very close to \dipsy. However, since the
elastic \pPb\ cross section comes out slightly larger than in \dipsy,
also the quasi-elastic cross section is larger.

The inelastic \pPb\ cross section has already been measured at LHC
\cite{CMS:2013rta,LHCb:2012aka}, and there is also a possibility to
measure the quasi-elastic cross section using \eg\ the TOTEM
experiment.  In table~\ref{tab:semiincxsec} we also present
predictions for the ratio between these cross sections, as one may
expect that some systematic uncertainties in the experimental
measurements (\eg\ uncertainties in luminosity) may cancel in such a
ratio.

For completeness we also show in table~\ref{tab:semiincxsec} the
\dipsy results for single diffractive excitation of the proton, which
is absent in all the Glauber calculations, and the double diffraction
cross section, which is zero in the black disc
approximations. Finally, we note that the ratio
$\sigma\subinelnd/\sigma\subtot$ by construction is one half in the
Glauber calculations (see \eqsref{eq:crossectionsGl} and
\eq{eq:greydisc}), while in \dipsy, the ratio is slightly higher.

%%%%%%%%%%%%%%%%%%%%%%%%%%%%%%%%%%%%%%%%%%%%%%%%%%%%%%%%%%%%%%%%%%%%%%%%%%%%%%%%%%%%%%%
%\clearpage

\section{Discussion}
\label{sec:effects}

\subsection{Saturation effects}
\label{sec:sat-eff}

Colour reconnection (swing) is one of the nuclear effects introduced in
the Lund Dipole Cascade Model \dipsy.  Such a swing may take place if
dipoles of the same colour are close enough in a longer chain of
dipoles, or, if the colour configuration of two different chains allows
for a colour reconnection.  This reconnection may happen with certain
probability discussed in \sectref{sec:dipsy} in such a way that
preferably smaller dipole chains are formed, possibly giving rise to
chains forming small closed loops.  Smaller dipoles reduce the
interaction probability and hence result in smaller cross sections.
This feature effectively corresponds to a gluon saturation effect, as
mentioned already at the description of the \dipsy model in
\sectref{sec:dipsy}.

In this section, we quantify the magnitude of the swing effect for
electron--nucleus (more precisely virtual photon--nucleus) and
proton--nucleus collisions. In particular, we switch on and off the
inter-nucleon swings. However, it is important to note that we keep
the \emph{intra}-nucleon swings turned on so that the values of the
total and elastic \pp\ cross sections remain the same as indicated in
\figref{fig:pp}.

It is well known that Gribov corrections to multiple diffractive
scattering~\cite{Gribov:1968jf} are typically on the $10-15$\% scale
and in principle they take into account the fluctuating size of the
projectile when evaluating, for example, the total cross section of
proton--nucleus or deuteron--nucleus collisions. In this picture, it
is important to consider that the projectile itself has a fluctuating
size, and when it is in a small size configuration, it can pass
through the nucleus easier. Hence the total \pA\ cross section is
typically reduced when the Gribov correction is taken into account.
Intra-nucleon swings may increase the fluctuations in the size of the
nucleon, and are hence related to these Gribov corrections. The
inter-nucleon swing mechanism is an additional, new effect, that may
decrease the probability of the interaction further, reducing the
impact of large colour dipole configurations even further.

In \figref{fig:eA-swing} the total cross sections with and without
inter-nucleon swing are compared for \gO, \gAu, \pO\ and \pPb\
reactions. As the number of possible swings will increase with the
number of dipoles, and hence with the collision energy, one might
assume that the effects of the inter-nucleon swing would also increase
with energy, however from \figref{fig:eA-swing} it is clear that the
energy dependence is practically flat. From this we can conclude that
the main effect comes from reconnections related to the initial
dipoles of the nucleons. For a virtual photon probe we find the
largest effect for a heavy nucleus (over 10\% for \gAu), which is
natural, since here we have more nucleons overlapping in
impact-parameter space. For a proton projectile, however, the effects
overall as well as the differences between nuclei are smaller, since
for central collisions (where the overlap is largest) at these
energies, the nucleus is already effectively ``black'', while the same
nucleus is much more transparent for a virtual photon probe.

\FIGURE[t]{
  \centering
  \epsfig{file=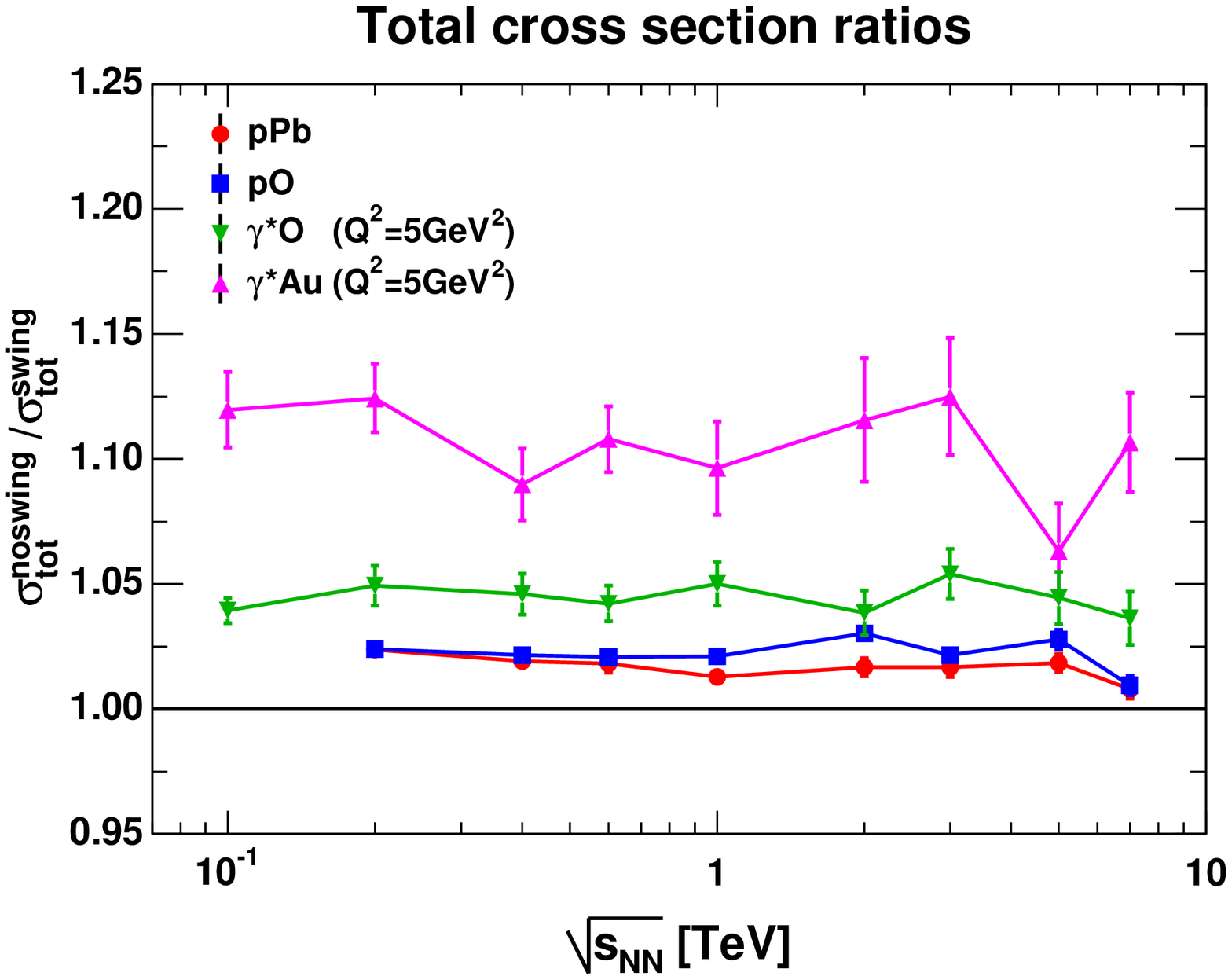,width=0.6\linewidth}
  \caption{Total cross section ratios for \pPb\ and \pO\ reactions, as
    well as for \gO\ and \gAu, with inter-nucleon of swing mechanism
    switched off as compared to on.  }
  \label{fig:eA-swing}
}

\subsection[Frame (in)dependence as a function of $\sigma^{\pp}\subtot$]
{\boldmath Frame (in)dependence as a function of $\sigma^{\pp}\subtot$}
\label{sec:frindep}

In this sub-section we demonstrate how can one minimise the dependence
of the results of the \dipsy model on the frame chosen as reference
frame for the simulation.  The reason for this discussion is that the
Lund Cascade Model \dipsy is not formulated in an entirely
boost-invariant manner, yet, so the simulation results depend slightly
on the Lorentz-frame chosen as a reference frame.

The results presented in \sectref{sec:results} are obtained using the
\NN\ (or \gN) centre-of-mass frame, with parameters tuned to \pp\ total
and elastic cross sections in this frame.
% (These parameters also
%reproduce the \gp\ cross sections.)
The screening effects in
interactions with nuclei are most sensitive to the cross section for
the individual \NN\ collisions.  Therefore, although the result
depends on the frame used when the \emph{parameter values are fixed},
as is shown in the left panel of \figref{fig:pApp-Y}, the result is
insensitive to the frame if the \emph{parameters are tuned, in the
  same frame, to \pp\ cross section data}.  This is illustrated on the
right panel of \figref{fig:pApp-Y}, which shows
$\sigma\subtot^{\pPb}/A\sigma\subtot^{\pp}$ as function of the \pp\
total cross section.

This way the error due to the frame dependence is reduced to a level
of about 1\% or less, as illustrated in the right panel of
\figref{fig:pApp-Y}, which negligible as compared to the $10-15$\%
size of the dynamical swing effect that we explored in the previous
sub-section.

\FIGURE[t]{
  \centering
  \epsfig{file=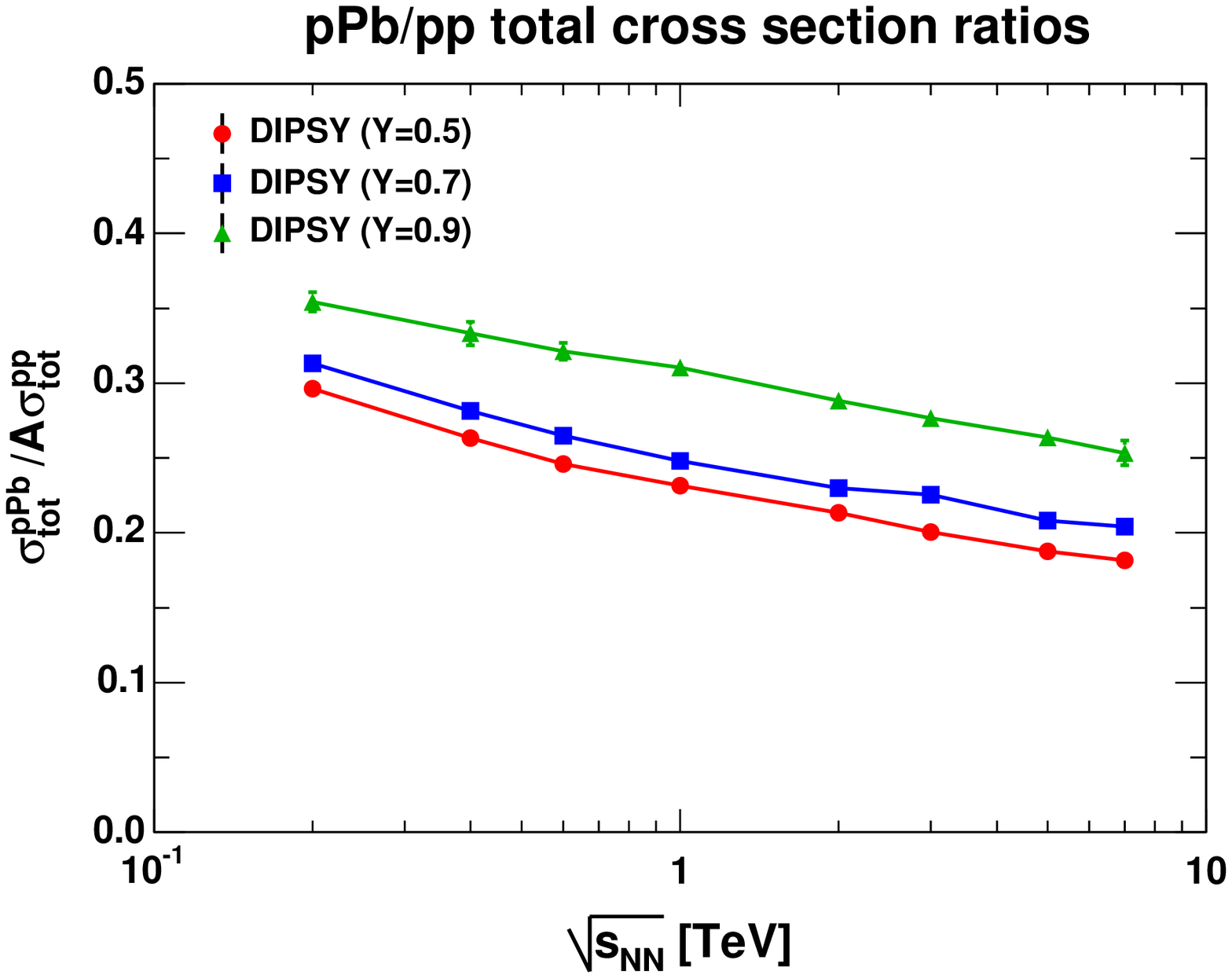,width=0.45\linewidth}
  \epsfig{file=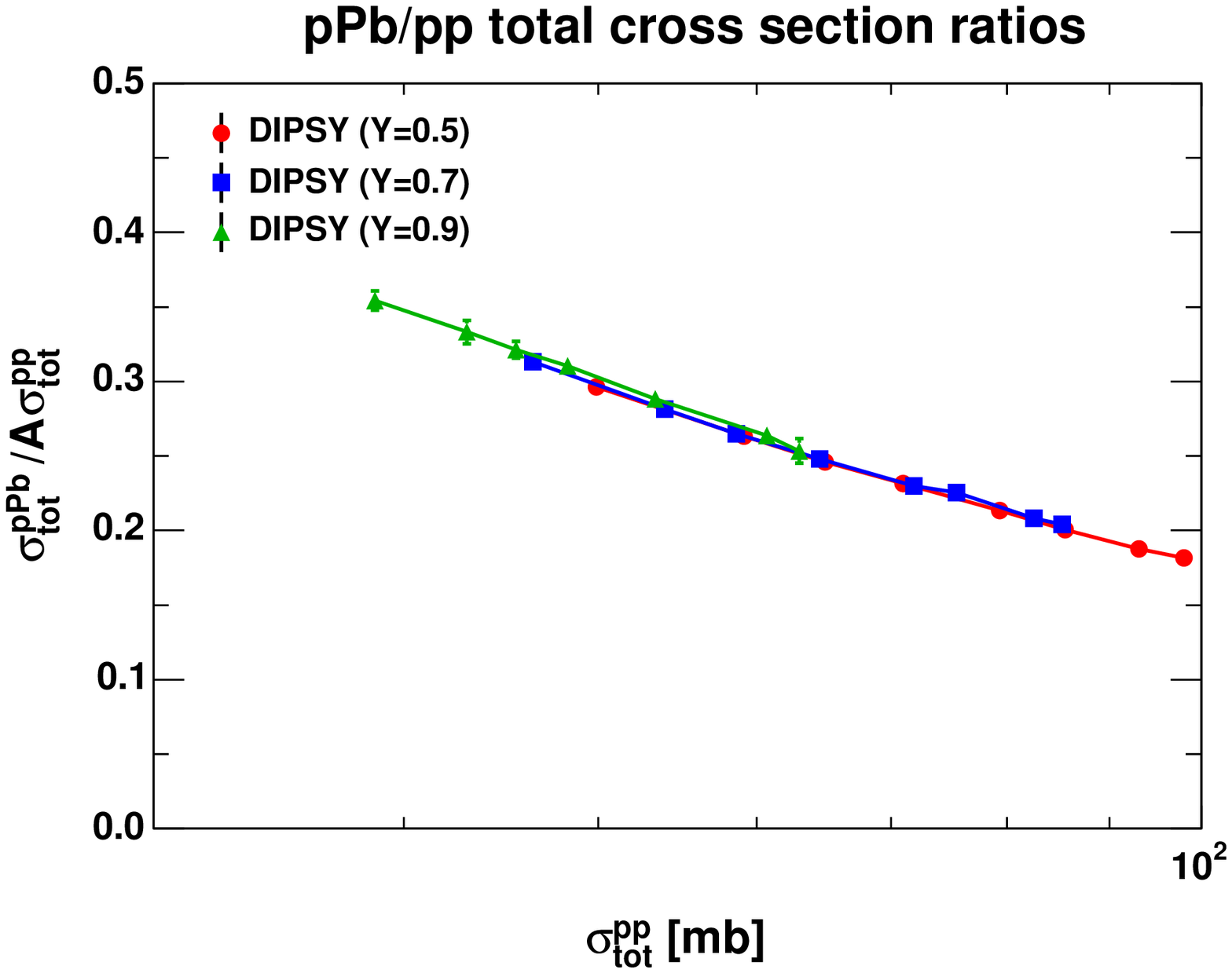,width=0.45\linewidth}
  \caption{ Left panel indicates the frame dependence of the
    normalised $\sigma\subtot^{\pPb}/A\sigma\subtot^{\pp}$ total cross
    section ratios calculated in three different frames,given by
    $y_{frame}=(1-Y)\cdot y_p + Y\cdot y_A$, with $Y=0.5,\,0.7$, and
    0.9, as a function of the centre-of-mass energy. We see that even
    these normalised ratios have a frame dependence of about 25\%. The
    right panel, however, illustrates that most of the frame
    dependence enters through the frame dependence of the total \pp\
    cross section: if these values are correctly tuned to the
    measurements in the frame of the calculation, the results (with
    less than 1\% error) become approximately frame independent. }
  \label{fig:pApp-Y}
}

\section{Conclusions}
\label{sec:conclusion}

Analyses of collective behaviour from final state properties in \AA\
collisions depend crucially on the assumptions about the initial
state, obtained from parton--parton interactions. This initial state
is sensitive to all possible correlations, fluctuations, and colour
interference effects between the partons within a nucleon as well as
in different nucleons. To draw stable conclusions, it is therefore
important to have a good understanding of this initial state.  It has
often been suggested that studies of \pA\ collisions, as intermediate
between \AA\ and \pp\ collisions, will be of great help here.

In this paper we will investigate the properties of a high energy
nucleus via studies of total, elastic, and quasi-elastic cross
sections in \pA\ and \gA\ collisions, as these observables depend only
upon the initial states, but are insensitive to any collective effects
in the final state interactions. The Lund Dipole Cascade model \dipsy
offers special possibilities to study the evolution of gluons inside
hadrons and nuclei, including effects of unitarization at small $x$,
interference, correlations, and fluctuations. The model is based on
BFKL evolution, including essential non-leading effects, saturation,
and colour interference effects. It reproduces successfully total,
elastic, and diffractive cross sections in \pp\ collisions and DIS,
within the energy range from RHIC to LHC for \pp\ collisions and the
HERA range for DIS, and it can be directly generalised to study
collisions with nuclei.

The general features of the results are, as expected, that in \pA\
interactions the centre is rather black, which means that the cross
section approximately scales with the area $(R_A+R_p)^2 \sim
R_p^2(A^{1/3}+1)^2$. In \gA\ collisions the nucleus is much more
transparent, in particular for high $Q^2$ and lighter nuclei. The
cross sections therefore here scale approximately with $A$, but
behaves somewhat more similar to \pA\ for smaller $Q^2$ and/or larger
$A$ and higher energy.  As a particularly interesting effect we note
that colour coherence effects between gluons in different nucleons in
the nucleus, can reduce the \gAu\ cross section by about 10\%.  These
results underline the importance of future electron--ion collider
experiments to study the initial partonic state of highly energetic
nuclei.

To study the more subtle effects related to colour coherence and
fluctuations, we make comparisons with Glauber calculations. We here
see that the interesting effects show up in the relative size of the
total, elastic, diffractive, and quasi-elastic cross sections. We
point out that both the experimental and the theoretical accuracy will
be higher for these ratios than for the individual cross sections.

% As a particularly interesting effect we also note that colour coherence
% effects between gluons in different nucleons in the nucleus, can
% reduce the \pPb\ cross section by about 10\%.

Finally we note that the \dipsy program is also able to produce
exclusive hadronic final states in \pp\ collisions
\cite{Flensburg:2011kk}. The extension of this to also model final
states of collisions involving heavy ions is certainly feasible, but
the effects of the extremely crowded partonic environment in \pA\ and
\AA\ are not trivial and must be modelled with great care. Progress
towards understanding string fragmentation in a dense environment has
already been made in \cite{Bierlich:2014xba}, but the full picture of
the final states in the \dipsy dipole model is yet to come.

\appendix

\section{Appendix. The Good--Walker formalism for diffractive excitation.} 
\label{sec:appendix}

Diffraction is normally interpreted as elastic scattering driven by
absorption.  If the absorption probability in Born approximation is
given by $2F$, then rescattering exponentiates in impact parameter
space, giving the unitarized absorption probability
\begin{equation}
  P_{abs}=d \sigma\subineltot/d^2b =1-e^{-2F}.
\end{equation}
We define the elastic amplitude, $T$, via the relation $S\equiv 1-T$,
which makes $T$ real. The optical theorem, $T=(1/2)(T^2+P_{abs})$,
then gives the following result for the amplitude and the elastic and
total cross sections:
\begin{eqnarray}
  T& = & 1-e^{-F}\\
  d\sigma\subel/d^2b&=& (1-e^{-F})^2\\
  d\sigma\subtot/d^2b&=& 2(1-e^{-F}).
\end{eqnarray}

For a projectile with a substructure, the mass eigenstates can differ
from the eigenstates of diffraction, \ie\ states with a specific
absorption probability~\cite{Good:1960ba}. Call the diffractive
eigenstates $\Phi_n$, with elastic scattering amplitudes $T_n$. The
mass eigenstates $\Psi_{k}$ are linear combinations of the states
$\Phi_n$:
\begin{equation}
  \Psi_{k} = \sum_n  c_{kn} \Phi_n\,\,\,\,\,\,(\mathrm{with}\,\,\Psi_{in}=\Psi_1).
  \label{eq:eigenstates}
\end{equation}
The elastic scattering amplitude is given by
\begin{equation}
  \langle \Psi_{1} | T | \Psi_{1} \rangle = \sum c_{1n}^2 T_n 
  = \langle T \rangle,
\end{equation}
and the elastic cross section
\begin{equation}
  d \sigma\subel/d^2 b = (\sum c_{1n}^2 T_n)^2 = \langle T\rangle ^2.
  \label{eq:sigmael}
\end{equation}
The amplitude for diffractive transition to the mass eigenstate
$\Psi_k$ is given by
\begin{equation}
  \langle \Psi_{k} | T | \Psi_{1} \rangle = \sum_n  c_{kn} T_n c_{1n},
\end{equation}
which gives a total diffractive cross section (including elastic
scattering)
\begin{equation}
  d\sigma\sub{diff}/d^2 b=\sum_k \langle \Psi_{1} | T | \Psi_{k} \rangle \langle
  \Psi_{k} | T | \Psi_{1} \rangle =\langle T^2 \rangle.
\end{equation}
Consequently the cross section for diffractive excitation is given by
the fluctuations:
\begin{equation}
  d\sigma\subd/d^2 b  = d\sigma\sub{diff}/d^2b- d \sigma\subel/d^2b =
  \langle T^2 \rangle - \langle T \rangle ^2.
  \label{eq:diffraction}
\end{equation}

If also the target has a substructure, it is possible to have either
single excitation of the projectile, of the target, or double
diffractive excitation, with cross sections given in
\sectref{sec:dipsy-model}.

\section*{Acknowledgments}
T. Cs. and A. Ster would like to thank L. and G. Gustafson and
L. L{\"o}nnblad for their kind hospitality during their visits to the
University of Lund, Sweden.  T. Cs. would also like to thank
R. J. Glauber for enlightening discussions and kind hospitality during
his visits to Harvard University.

\bibliographystyle{utcaps}  
\bibliography{xsect} 

\end{document}